\documentclass[12pt]{article}
\pdfoutput=1
\usepackage{amsmath}
\usepackage{amssymb}
\usepackage{graphicx}
\usepackage{subfigure}
\usepackage{url}
\usepackage{comment}
\usepackage[sort&compress, numbers, merge]{natbib}

\newcommand{\Slash}[1]{{\ooalign{\hfil/\hfil\crcr$#1$}}}

\usepackage[colorlinks=true, linkcolor=blue, citecolor=blue,
urlcolor=black]{hyperref}

\begin{document}
\baselineskip 0.6cm

\def\simgt{\mathrel{\lower2.5pt\vbox{\lineskip=0pt\baselineskip=0pt
           \hbox{$>$}\hbox{$\sim$}}}}
\def\simlt{\mathrel{\lower2.5pt\vbox{\lineskip=0pt\baselineskip=0pt
           \hbox{$<$}\hbox{$\sim$}}}}
\def\simprop{\mathrel{\lower3.0pt\vbox{\lineskip=1.0pt\baselineskip=0pt
             \hbox{$\propto$}\hbox{$\sim$}}}}
\def\tr{\mathop{\rm tr}}

\begin{titlepage}

\begin{flushright}
DESY 14-180 \\
FTPI-MINN-14/37\\
IPMU14-0320
\end{flushright}

\vskip 1.1cm

\begin{center}

{\large \bf 
Higgsino Dark Matter in High-Scale Supersymmetry
}

\vskip 1.2cm

Natsumi Nagata${}^{1}$ and
Satoshi Shirai${}^{2}$
\vskip 0.4cm

{\it
$^1$ William I. Fine Theoretical Physics Institute, School of
Physics and Astronomy, \\ 
University of Minnesota, Minneapolis, MN 55455, USA,\\ 
and Kavli Institute for the Physics and Mathematics of the
  Universe (WPI), 
  \\Todai Institutes for Advanced Study, University of Tokyo, Kashiwa
 277-8583, Japan\\ [5pt]
$^2$ {Deutsches Elektronen-Synchrotron (DESY), 22607 Hamburg, Germany}
}

\vskip 0.8cm

\abstract{

 We study a supersymmetric (SUSY) Standard Model in which a Higgsino is
 light enough to be dark matter, while the other SUSY particles are much
 heavier than the weak scale. We carefully treat the effects of
 heavy SUSY particles to the Higgsino nature, especially taking into
 account the renormalization effects due to the large hierarchy between
 the Higgsino and the SUSY breaking scales. Inelastic scattering of the
 Higgsino dark matter with a nucleus is studied, and the constraints on
 the scattering by the direct detection experiments are discussed. This
 gives an upper limit on the new physics scale. Bounds on the dark
 matter-nucleon elastic scattering, the electric dipole moments, and
 direct production of Higgsinos, on the other hand, give a lower
 limit. We show the current status on the limits and discuss the future
 prospects. 

}

\end{center}
\end{titlepage}

\section{Introduction}
\label{sec:intro}

The supersymmetric (SUSY) Standard Model (SSM) is a strong candidate for
new physics. The weak-scale SUSY is commonly said to provide a solution
to the hierarchy problems, promising frameworks for the grand
unification, and the correct amount of dark matter (DM) in the
Universe. However, the discovery of the Standard Model (SM)-like Higgs
boson with a mass of $\sim 125$~GeV~\cite{Aad:2012tfa,Chatrchyan:2012ufa}
as well as the absence of new physics seems to imply the SUSY breaking
scale is much higher than the weak scale~\cite{Wells:2003tf,*Wells:2004di,
ArkaniHamed:2004fb,*Giudice:2004tc,*ArkaniHamed:2004yi,*ArkaniHamed:2005yv}. 
With the SUSY breaking scale larger than $O(10)$ TeV, the observed Higgs
mass can be realized
\cite{Okada:1990vk,*Okada:1990gg,*Ellis:1990nz,*Haber:1990aw,
*Ellis:1991zd,Giudice:2011cg}. The high-scale SUSY scenario may offer an
even more precise gauge coupling unification~\cite{Hisano:2013cqa} and
open up possibilities for the simplest framework of the grand unified
theory~\cite{Hisano:2013exa,Nagata:2013sba}. With the $R$-parity
conservation assumed, it also provides the lightest SUSY particle (LSP)
as a DM candidate. In addition, such a high-scale SUSY scenario can 
greatly relax serious SUSY flavor/CP~\cite{Gabbiani:1996hi,
Moroi:2013sfa, McKeen:2013dma, Sato:2013bta, Altmannshofer:2013lfa,
Fuyuto:2013gla} and cosmological problems~\cite{Weinberg:1982zq,
Kawasaki:2004fw, Kawasaki:2008qe}. For these reasons, this framework has
been gathering 
more and more attention these days, especially after the discovery of
the Higgs boson~\cite{Hall:2011jd, Hall:2012zp, Ibe:2011aa,
*Ibe:2012hu, Arvanitaki:2012ps, ArkaniHamed:2012gw}.

In such a high scale SUSY model, however, the weak scale can be realized
only with a great extent of fine-tuning. Although the origin of stability of
the weak scale is unclear, an appealing approach would be utilizing the
anthropic principle or environmental selection on multiverse; the
$O(100)$ GeV weak scale is essential for the formulation of complex
nuclei~\cite{Agrawal:1997gf} that is crucial for the existence of
intelligent life, just as in the case of the cosmological
constant~\cite{Weinberg:1987dv}. 

This kind of environmental selection may also work on the LSP
mass $m_{\text{LSP}}
$ \cite{Elor:2009jp,Hall:2011jd,Hall:2012zp,Hall:2013eko,
Nomura:2014asa}. A too heavy LSP mass leads to over-abundance of DM in  
the Universe. To avoid this catastrophe, the LSP mass should be
significantly tuned to be around TeV scale or much heavier than the mass
scale of inflaton. If too much abundance of DM is disfavored with the
environmental selection~\cite{Tegmark:2005dy}, a mass region
\begin{align}
O(1-10)~{\rm TeV} \simlt m_{\rm LSP} \simlt \max\{10^2 ~T_R, m_{\rm inf}\},
\end{align}
may be forbidden, where $T_R$ is the reheating temperature of the
Universe and $m_{\rm inf}$ is the inflaton mass~\cite{Hall:2014vga}. The
recent report on the search for gravitational waves by
BICEP2~\cite{Ade:2014xna}, for instance, may indicate $m_{\rm inf} \sim
10^{13}$ GeV, though the interpretation of the result is controversial
\cite{Adam:2014oea}. Further, $T_R\simeq 10^9$~GeV is necessary
condition for the successful thermal leptogenesis
\cite{Fukugita:1986hr}. Anyway, we expect large hierarchy between the
electroweak scale and the energy scale of $T_R$. To evade the above
unacceptable window, an environmental selection may work to let the LSP
mass remain TeV scale, which results in a considerable fine-tuning for
the LSP mass parameter. In this case, the ``lonely LSP'' scenario, in
which only the LSP is around TeV scale and the other SUSY particles are
much heavier, can be realized. 

Even without such an anthropic viewpoint, the ``lonely LSP'' scenario
can be achieved for some dynamical reasons. For example, if a certain
symmetry forbids the tree-level LSP mass and it is generated only by
radiative corrections, the LSP mass will be much suppressed compared to
those of the other SUSY particles. Among the minimal SSM (MSSM)
particles, an experimentally viable candidate for the LSP DM is a
Higgsino or a Wino.
Although a Bino or a gravitino LSP would be possible,
its abundance strongly depends on the high-energy model and tends to be
produced too much. 
The Wino DM case has been widely
considered so far \cite{Gherghetta:1999sw, Moroi:1999zb} since it is
motivated by the anomaly mediation \cite{Randall:1998uk,
Giudice:1998xp}, and their phenomenology is thoroughly discussed in
previous works \cite{Hisano:2003ec, *Hisano:2004ds, *Hisano:2006nn,
ArkaniHamed:2006mb, Shirai:2009fq, *Ibe:2013jya, *Ibe:2014qya,
Hisano:2010fy, *Hisano:2010ct,
Hall:2012zp, Ibe:2012sx, Cohen:2013ama, *Fan:2013faa,
Bhattacherjee:2014dya}. The Higgsino LSP is also viable, for its mass
can be suppressed by some symmetries such as the Peccei-Quinn symmetry
\cite{Peccei:1977hh} or the $R$-symmetry. In this paper, we focus on
this Higgsino LSP case. Indeed, the Higgsino mass with a mass
of $\sim 1$~TeV can explain the observed DM density
\cite{Cirelli:2007xd}, while the environmental selection arguments
may suggest that the Higgsino LSP has a mass of $\mathcal{O}(100)$~GeV
(unless it is much heavier than the inflation scale). This mass region
is the target of the present study. For the former arguments, see,
\textit{e.g.}, Refs.~\cite{Cheung:2005pv, Cheung:2009qk, Beylin:2009wz,
Jeong:2011sg, Fox:2014moa}.

The ``lonely Higgsino'' actually cannot be completely lonely, for
a pure Higgsino DM has been already excluded by the DM direct detection
experiments. Tiny amount of mixing among the Higgsino and gauginos is
required to avoid the constraints, which gives an upper-bound on the
SUSY breaking scale. It turns out that the scale is much larger than the
TeV scale. Such a large mass hierarchy induces large quantum
corrections. Thus, to study the properties of the Higgsino DM
precisely, we need to take the effects into account.

In this work, we revisit the phenomenology of the Higgsino LSP
considering the renormalization corrections due to the large hierarchy
between the Higgsino mass and the SUSY breaking scales. These
corrections affect the mass splitting between the neutral Higgsinos,
which are important to discuss the constraints on it coming from the
inelastic scatterings of the Higgsino DM with a nucleon. We will study these
constraints in the case of the Higgsino DM in detail and by using the
results derive an upper limit on the gaugino mass scale. The mass splitting
depends on new CP-phases appearing in the gaugino and Higgsino masses as
well, and the phases can be probed by means of the electric dipole moments
(EDMs). We will discuss the interplay between the bounds from the
EDM measurements and the DM direct detection experiments. The elastic
scattering of the Higgsino DM with a nucleon, as well as the direct
production of Higgsinos in colliders, is also discussed with their future
prospects. We will find that the constraints from the measurements of the
above quantities are complementary to each other. By considering them
altogether, we may probe the nature of the Higgsino DM and the signature
of high-scale physics in future experiments, which enables us to gain an
insight on the SSM.

The organization of this paper is as follows. In the next section, we
study the mass spectrum of Higgsinos and new physics effects on it. The
effects are expressed in terms of the dimension-five effective
operators. Then, in Sec.~\ref{sec:RGEhigherdim}, we present the
renormalization group equations (RGEs) for the operators as well as
their matching conditions, and study the renormalization effects on
them. By using the results, we discuss the constraints on the Higgsino
DM scenario from the direct detection experiments, the measurements of
the EDMs, and the Higgsino searches in colliders in
Sec.~\ref{sec:DMsearch}, Sec.~\ref{sec:EDM}, and
Sec.~\ref{sec:collider}, respectively. Section~\ref{sec:summary} is
devoted to summary of the results and discussion.

\section{Higgsino Mass Spectrum}

To begin with, we give a brief review on the mass spectrum of Higgsinos
in the presence of small mixing with gauginos whose masses are assumed
to be much heavier than the Higgsino masses. The dominant mixing effects are
included in the dimension-five effective operators shown below. Their
coefficients as well as the renormalization effects on them are evaluated
in the subsequent section.

In the MSSM, the mass term for Higgsinos $\widetilde{H}_u$ and
$\widetilde{H}_d$ is given as
\begin{equation}
 {\cal L}_{\rm Higgsino~mass}=
-\mu ~\epsilon^{\alpha\beta}
(\widetilde{H}_{u}^{})_\alpha^{} (\widetilde{H}^{}_{d})_\beta^{} 
+{\rm h.c.}~,
\end{equation}
where $\alpha$ and $\beta$ are the SU(2)$_L$ indices,
$\epsilon^{\alpha\beta}$ is an antisymmetric tensor with
$\epsilon^{12}=- \epsilon^{21}=+1$, and 
\begin{equation}
 \widetilde{H}^{}_u=
\begin{pmatrix}
 \widetilde{H}_u^+\\ \widetilde{H}_u^0
\end{pmatrix}
~,~~~~~~
 \widetilde{H}^{}_d=
\begin{pmatrix}
 \widetilde{H}_d^0\\ \widetilde{H}_d^-
\end{pmatrix}
~.
\end{equation}
As one can see, $\widetilde{H}_u$ and $\widetilde{H}_d$ form a Dirac
fermion. Thus, there is a U(1) symmetry under which $\widetilde{H}_u$
and $\widetilde{H}_d$ are oppositely charged. If there exist
operators which break the U(1) symmetry, however, the Dirac fermion is
divided into a pair of Majorana fermions. Up to dimension-five, such
operators are given as\footnote{Notice that operators like
$\epsilon^{\alpha\beta}\epsilon^{\gamma\delta}  (H)_\alpha (H)_\beta
(\widetilde{H}_d)_\gamma  (\widetilde{H}_d)_\delta$ vanish since the
Higgs field is bosonic. } 
\begin{equation}
{\cal L}_{\rm eff}= \sum_{i=1,2}c_i \mathcal{O}_i +{\rm h.c.}~,
\label{eq:effc}
\end{equation}
where
\begin{align}
 \mathcal{O}_1&\equiv
(H^\dagger)^\alpha (\widetilde{H}_u)_\alpha 
(H^\dagger)^\beta (\widetilde{H}_u)_\beta ~, \nonumber \\
 \mathcal{O}_2 &\equiv
\epsilon^{\alpha\beta}\epsilon^{\gamma\delta} 
(H)_\alpha (\widetilde{H}_d)_\beta
(H)_\gamma (\widetilde{H}_d)_\delta ~,
\end{align}
and
\begin{equation}
 H=
\begin{pmatrix}
 H^+\\ H^0
\end{pmatrix}
\end{equation}
denotes the SM Higgs field. These operators give rise to the mass
splitting between the neutral components of the Higgsinos. We also have the
dimension-five operators that do not violate the U(1) symmetry:
\begin{equation}
  {\cal L}_{\rm eff}=\sum_{i=1,2}d_i\widetilde{\mathcal{O}}_i
+\text{h.c.}~,
\end{equation}
with
\begin{align}
\widetilde{\mathcal{O}}_1&\equiv
 \epsilon^{\beta\gamma}(H^\dagger)^\alpha
 (\widetilde{H}_{u}^{})_\alpha^{}  
(H)_\beta^{} (\widetilde{H}^{}_{d})_\gamma^{} 
~, \nonumber \\
\widetilde{\mathcal{O}}_2&\equiv
 \epsilon^{\beta\gamma}(H^\dagger)^\alpha
 (\widetilde{H}_{d}^{})_\alpha^{}  
(H)_\beta^{} (\widetilde{H}^{}_{u})_\gamma^{} ~.
\label{eq:effd}
\end{align}
These two operators yield the
mass difference between the neutral and charged components. 
Note that by using the Fierz identities one can easily show that
\begin{equation}
 \epsilon^{\alpha\beta}|H|^2 
(\widetilde{H}_{u}^{})_\alpha^{} (\widetilde{H}^{}_{d})_\beta^{}=
\widetilde{\mathcal{O}}_1-\widetilde{\mathcal{O}}_2~.
\end{equation}
Therefore, the operators $\mathcal{O}_i$ and $\widetilde{\mathcal{O}}_i$
exhaust the dimension-five operators which consist of the Higgsinos and
the Higgs field and are allowed by the gauge and Lorentz symmetries. 

Let us examine the mass differences induced by the above
operators. After the electroweak symmetry breaking, the mass matrix for
the neutral components is given by  
\begin{equation}
 {\cal L}_{\rm mass}=
-\frac{1}{2}(\widetilde{H}^0_d ~\widetilde{H}^0_u)
{\cal M}
\begin{pmatrix}
 \widetilde{H}^0_d \\ \widetilde{H}^0_u
\end{pmatrix}
+{\rm h.c.}~,
\end{equation}
with
\begin{equation}
 {\cal M}=
\begin{pmatrix}
 -{v}^2(|\mu|) c_2(|\mu|)& -\bar{\mu}\\
-\bar{\mu} & -{v}^2(|\mu|)c_1(|\mu|)
\end{pmatrix}
~,
\end{equation}
where $v\simeq 246$~GeV is the Higgs vacuum expectation value (VEV) and
$\bar{\mu}\equiv \mu 
-{{v}^2(|\mu|)}d_1(|\mu|)/2$. The parameters and the Wilson coefficients
in the mass matrix are renormalized at the scale of $|\mu|$. We omit the
argument in the following discussion, for
brevity. The mass matrix ${\cal M}$ is diagonalized\footnote{In Appendix
~\ref{sec:takagi}, we summarize formulae for the diagonalization
of a $2\times 2$ complex symmetric matrix. } by using an unitary
matrix $N$ as
\begin{equation}
 N^*{\cal M}N^\dagger=
\begin{pmatrix}
 m_1&0\\ 0 & m_2
\end{pmatrix}
~,
\end{equation}
and the resultant masses $m_1$ and $m_2$ are given as
\begin{align}
 m_1&\simeq |\bar{\mu}|-\frac{|{\mu}^*c_1 +{\mu} c_2^*|}
{2|{\mu}|}{v}^2~,\\
 m_2&\simeq |\bar{\mu}|+\frac{|{\mu}^*c_1 +{\mu} c_2^*|}
{2|{\mu}|}{v}^2~,
\end{align}
where we keep the ${\cal O}({v}^2)$ terms. 
In this case, the mass difference between the neutral components is
found to be\footnote{The result differs from that presented in
Ref.~\cite{Essig:2007az}. }
\begin{equation}
 \Delta m \equiv m_2-m_1 \simeq \frac{|{\mu}^*c_1 +{\mu} c_2^*|}
{|{\mu}|}{v}^2~.
\label{eq:neutmassdiff}
\end{equation}
The expression indicates that the mass difference depends on the phases
in the $\mu$-term and the Wilson coefficients $c_1$ and $c_2$. The
unitary matrix $N$ is evaluated as 
\begin{equation}
 N= e^{\frac{i}{2}\phi_\mu}
\begin{pmatrix}
 e^{-\frac{i}{2}(\phi+\alpha)} \cos\theta  
& -e^{\frac{i}{2}(\phi-\alpha)}\sin\theta \\
ie^{-\frac{i}{2}(\phi+\beta)} \sin\theta & ie^{\frac{i}{2}(\phi
-\beta)} \cos\theta 
\end{pmatrix}
~,
\end{equation}
with
\begin{equation}
 \tan \theta \simeq 1+\frac{(|c_2|^2-|c_1|^2)v^2}{2|{\mu}^*c_1 +{\mu}
  c_2^*| }~,
\end{equation}
and
\begin{align}
 \phi &= \arg(\bar{\mu}^*c_1 +\bar{\mu} c_2^*)~,
~~~~~~\phi_\mu =\arg({\mu})~~, \nonumber \\[3pt]
 \alpha &=\frac{v^2}{2}\text{Im}\biggl(
\frac{d_1+2c_2e^{i\phi}}{\mu}\biggr)~,
~~~~~~
\beta=\frac{v^2}{2}\text{Im}\biggl(
\frac{d_1-2c_1e^{-i\phi}}{\mu}
\biggr)~.
\end{align}
Again, we remain the terms up to $\mathcal{O}(v^2)$. 
By using the unitary matrix, the mass eigenstates are written as follows: 
\begin{equation}
\begin{pmatrix}
 \widetilde{\chi}^0_1 \\ \widetilde{\chi}^0_2
\end{pmatrix}
=N
 \begin{pmatrix}
  \widetilde{H}^0_d \\ \widetilde{H}^0_u
 \end{pmatrix}
~.
\end{equation}
Here, $\widetilde{\chi}^0_1$ and $\widetilde{\chi}^0_2$ are the mass
eigenstates corresponding to $m_1$ and $m_2$, respectively. 

The mass term of the charged Higgsino is, on the other hand, given by
\begin{equation}
 {\cal L}_{\rm mass}=-(\mu +\frac{v^2}{2}d_2)\widetilde{H}^+_u 
\widetilde{H}^-_d + {\rm h.c.}~.
\end{equation}
Through the field redefinition, we can write the mass term with the mass
eigenstate $\widetilde{\chi}^+$ as
\begin{equation}
 {\cal L}_{\rm mass}=-m_{\widetilde{\chi}^\pm}\overline{\widetilde{\chi}^+}
\widetilde{\chi}^+ + {\rm h.c.}~.
\end{equation}
Here, $\widetilde{\chi}^+$ is a four-component Dirac fermion defined by
\begin{equation}
 \widetilde{\chi}^+\equiv
\begin{pmatrix}
 e^{\frac{i}{2}(\phi_\mu+\gamma)} \widetilde{H}^+_u \\  
 e^{-\frac{i}{2}(\phi_\mu +\gamma)}(\widetilde{H}^-_d)^\dagger
\end{pmatrix}
~,
\end{equation}
with 
\begin{equation}
 m_{\widetilde{\chi}^\pm}= |\mu +\frac{v^2}{2}d_2|~,
 ~~~~~~\gamma =\frac{v^2}{2}\text{Im}\biggl(\frac{d_2}{\mu}\biggr)~.
\end{equation}
From the mass parameters obtained above, one can easily find that the
higher-dimensional operators also contribute to the mass difference between
charged Higgsino and the Higgsino DM. The contribution $\Delta
m_+|_{\text{tree}}$ is given by
\begin{equation}
\Delta m_+|_{\text{tree}}
\simeq \frac{v^2}{2}\biggl[
|\mu|{\rm Re}\biggl(
\frac{d_1+d_2}{\mu}
\biggr)
+\frac{|{\mu}^*c_1 +{\mu} c_2^*|}{|\mu|}
\biggr]~.
\end{equation}

\begin{figure}[t!]
\begin{center}
 \includegraphics[clip, width = 0.5 \textwidth]{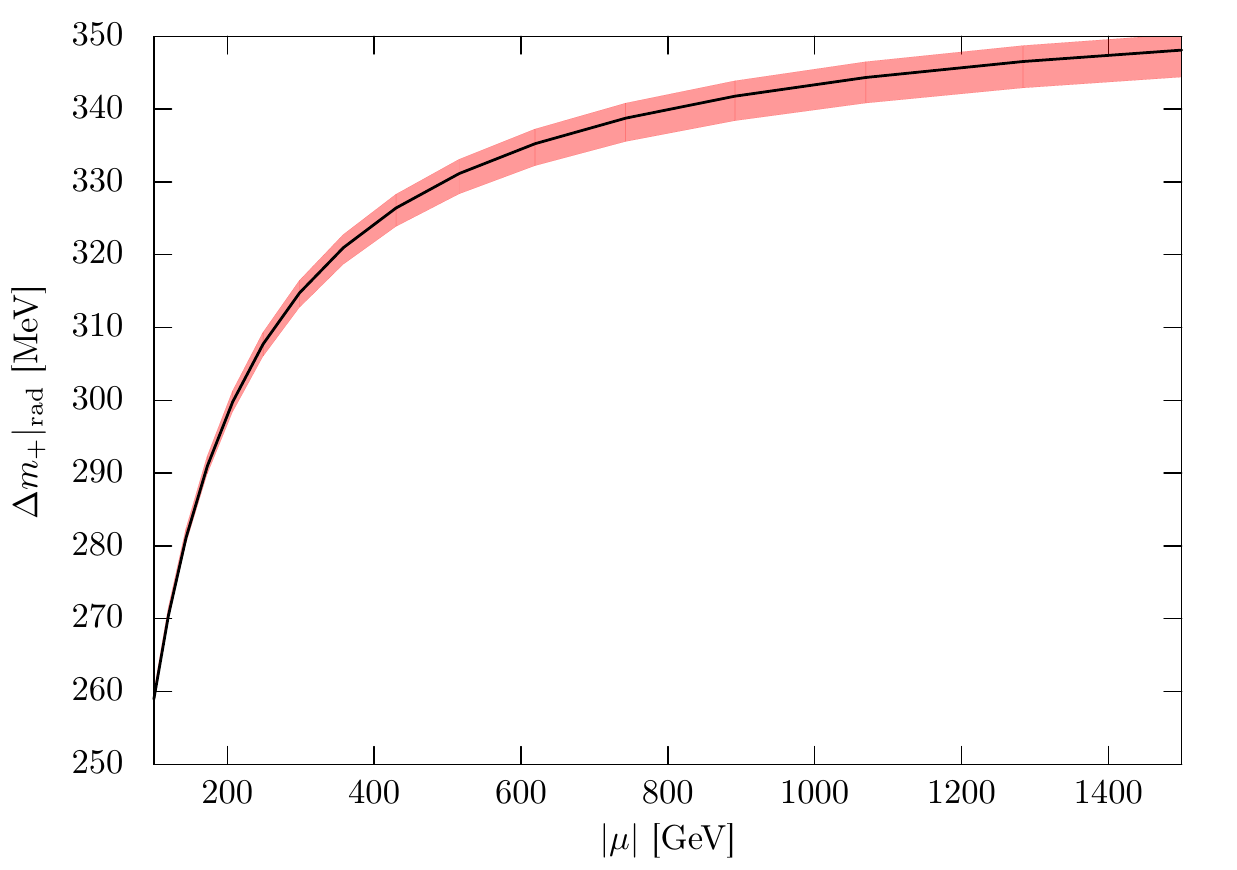}
\caption{
The radiative corrections to the neutral-charged Higgsino
mass difference $\Delta m_+|_{\text{rad}}$ as a function of the Higgsino
 mass parameter $|\mu|$. Red band represents uncertainty coming from the
 higher-loop contribution. }
\label{fig:cnrad}
\end{center}
\end{figure}

In addition, it is known that radiative corrections by the electroweak
gauge bosons induce the neutral-charged Higgsino mass difference. At
one-loop level, the contribution is expressed as
\begin{equation}
\Delta m_+|_{\text{rad}} =\frac{\alpha_2}{4\pi}m_{\widetilde{\chi}^\pm}
\sin^2\theta_W f
\biggl(\frac{m_Z}{m_{\widetilde{\chi}^\pm}}\biggr)~,
\label{eq:cnrad}
\end{equation}
where $\alpha_2\equiv g^2/(4\pi)$ with $g$ the SU(2)$_L$ gauge coupling
constant, $\theta_W$ is the weak mixing angle, and $m_Z$ is the mass of
$Z$ boson.  The function $f(x)$ is given by\footnote{
We also give an analytic expression of $f(x)$:
\begin{equation}
f(x)=-x^2+x^4\ln (x) +4x\biggl(1+\frac{x^2}{2}\biggr)\sqrt{1-\frac{x^2}{4}}
\tan^{-1}\biggl(\frac{2}{x}\sqrt{1-\frac{x^2}{4}}\biggr)~.
\end{equation}
}
\begin{equation}
 f(x)=2\int^1_0 dt~(1+t)\ln
\biggl(1+\frac{x^2(1-t)}{t^2}\biggr)~.
\end{equation}
Especially, in the limit of $x\to 0$, 
\begin{equation}
f(x)\simeq 2\pi x -3x^2  +\dots ~,
\end{equation}
and thus Eq.~\eqref{eq:cnrad} is approximated by
\begin{equation}
 \Delta m_+|_{\text{rad}}
\simeq  \frac{1}{2}\alpha_2 m_Z \sin^2\theta_W \biggl(1-\frac{3m_Z}
{2\pi m_{\widetilde{\chi}^\pm}}\biggr)~.
\end{equation}
In Fig.~\ref{fig:cnrad}, we show the radiative corrections to the
neutral-charged Higgsino mass difference $\Delta m_+|_{\text{rad}}$ as
a function of the Higgsino mass parameter $|\mu|$. Here, the red band
represents  uncertainty coming from the higher-loop contribution.
We will see below that the radiative correction is comparable or even
dominates the contribution of the higher-dimensional operators $\Delta
m_+|_{\text{tree}}$ in a wide range of parameter region. 

After all, the mass difference between the neutral and charged
components is given by
\begin{equation}
\Delta m_+ \equiv
 m_{\widetilde{\chi}^\pm}-m_{\widetilde{\chi}^0}
=\Delta m_+|_{\text{tree}}+\Delta m_+|_{\text{rad}}~,
\end{equation}
where we define $m_{\widetilde{\chi}^0}\equiv m_1$.
It plays an important role when we study the collider phenomenology of
Higgsinos, as discussed in Sec.~\ref{sec:collider}.

In the following analysis, we use the above resummed dimension five
operators for estimations of low-energy observables. 
As for contributions which cannot be covered only with the dimension
five operators, we use the tree level result.

\section{Renormalization of Higher Dimensional Operators}
\label{sec:RGEhigherdim}

The dimension-five effective operators discussed above are induced by
the Bino and Wino exchanging processes at the gaugino mass scale. Let us
evaluate the matching conditions. First, we present our
convention for the definition of the gaugino masses and the
gaugino-Higgsino-Higgs couplings. The gaugino mass terms are defined by
\begin{align}
{\cal L}_{\text{gaugino mass}}=
-\frac{M_1}{2}\widetilde{B}\widetilde{B}
-\frac{M_2}{2}\widetilde{W}^a \widetilde{W}^a +\text{h.c.}~,
\end{align}
where $\widetilde{B}$ and $\widetilde{W}^a$ represent Bino and Wino,
respectively, with $a$ being the SU(2)$_L$ adjoint index. Relevant
Yukawa interactions of the Higgs boson, Higgsinos and gauginos are given by
\begin{align}
{\cal L}_{\text{int}} =  -\frac{1}{\sqrt{2}}
\{g_{1u} H^{\dagger} \widetilde{H}_u+
 g_{1d} \epsilon^{\alpha\beta} (H)_\alpha(\widetilde {H}_d)_\beta
 \} \widetilde{B} 
  -
  \frac{1}{\sqrt{2}}\{
g_{2u} H^{\dagger} \sigma^a \tilde H_u 
 -g_{2d} \epsilon^{\alpha\beta}(H)_\alpha (\sigma^a \widetilde{H}_d)_\beta
  \}\widetilde{W}^a +\text{h.c.}~,
\label{eq:gauginohiggsinocoup}
\end{align}
where $\sigma^a$ are the Pauli matrices, and the above couplings at
leading order are given as
\begin{align}
g_{1u} &= g^\prime \sin\beta, ~~~~~~g_{1d} = g^\prime \cos\beta~, \nonumber\\
g_{2u} &= g \sin\beta, ~~~~~~~g_{2d} = g \cos\beta~,
\label{eq:gauginocoupmatch}
\end{align}
at the SUSY breaking scale. Here, $g^\prime$ is the U(1)$_Y$ gauge
coupling constant, and $\tan\beta
\equiv \langle H_u^0\rangle /\langle H_d^0\rangle$. 
Then, by integrating out the gauginos, we obtain the matching conditions
for the Wilson coefficients at the gaugino mass scale as follows:
\begin{align}
c_{1} &= \frac{g_{1u}^2}{4M_1}  + \frac{g_{2u}^2}{4M_2}~, \nonumber \\
c_{2} &= \frac{g_{1d}^2}{4M_1}  + \frac{g_{2d}^2}{4M_2}~, \nonumber \\ 
d_{1} &= \frac{g_{1u} g_{1d}}{2M_1}  +  \frac{g_{2u} g_{2d}}{2M_2}~,
 \nonumber \\  
d_{2} &= -\frac{g_{2u} g_{2d}}{M_2}~,  \label{eq:BC}
\end{align}
with all of the parameters determined at the gaugino mass scale.

\begin{figure}[t!]
\begin{center}
 \includegraphics[clip, width = 0.8 \textwidth]{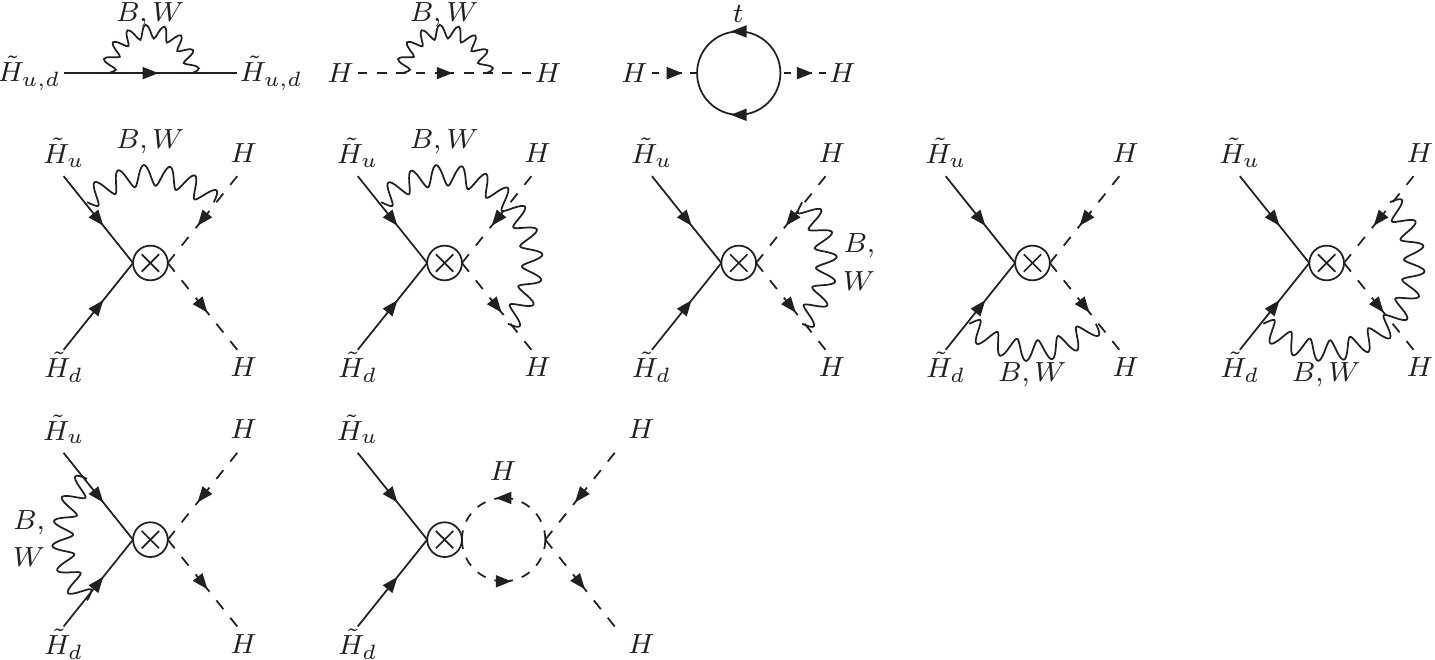}
\caption{
Examples of diagrams relevant for the RGEs.
}
\label{fig:RG}
\end{center}
\end{figure}

These Wilson coefficients are evolved down to the Higgsino mass scale
according to the RGEs which we obtain by computing the diagrams in
Fig.~\ref{fig:RG}:\footnote{
The RGE~\eqref{Eq:RGEc} can be read from that for the dimension-five
operator for the neutrino masses \cite{Chankowski:1993tx, Babu:1993qv}. 
The RGEs for other coupling constants are presented in
Appendix~\ref{sec:RGEs}. }
\begin{align}
 \frac{d c_i}{d\ln Q}
=\frac{1}{16\pi^2}(6y_t^2+2\lambda -3g^2)c_i~,
\label{Eq:RGEc}
\end{align}
for $i=1,2$, and  
\begin{equation}
 \frac{d }{d\ln Q}(d_1, d_2)
=(d_1, d_2)\cdot \frac{1}{16\pi^2}
\begin{pmatrix}
 6y_t^2+4\lambda-3g^{\prime 2}-6g^2& -2\lambda +3g^2 \\
-2\lambda +3g^2 & 6y_t^2+4\lambda-3g^{\prime 2}-6g^2
\end{pmatrix}
~.
\end{equation}
Here, $y_t$ is the top Yukawa coupling and $\lambda$ is the Higgs
self-coupling given by 
\begin{equation}
 {\cal L}_{\rm self}=-\frac{\lambda}{2} (|H|^2)^2~,
\end{equation}
and we neglect the other Yukawa couplings than that of top quark.

To see the significance of the renormalization effects,
as an example, we consider the case where the higher dimensional
operators dominantly arise from the Wino exchange. At tree level, we have
\begin{align}
c_1 v^2|_{\rm tree} &= \frac{m_W^2 \sin^2\beta}{M_2}~, ~~~~~~~~~~
c_2 v^2|_{\rm tree} = \frac{m_W^2 \cos^2\beta}{M_2}~, \nonumber\\
d_1 v^2|_{\rm tree} &= \frac{2m_W^2 \sin\beta \cos\beta}{M_2}~, ~~~
d_2 v^2|_{\rm tree} = \frac{-4 m_W^2 \sin\beta \cos\beta}{M_2}~,
\end{align}
with $m_W$ the $W$-boson mass. 
Let us define the ratio of the renormalized values to the tree level
values, $R_{c_i}$ and $R_{d_i}$ $(i=1,2)$ such that
\begin{align}
R_{c_i} \equiv \frac{c_i(|\mu|) v^2(|\mu|)}{c_i v^2|_{\rm tree}}~,~~~~~~
R_{d_i} \equiv \frac{d_i(|\mu|) v^2(|\mu|)}{d_i v^2|_{\rm tree}}~.
\end{align}
Here we evaluate the running Higgs VEV $v$ according to
Ref.~\cite{Pierce:1996zz} as 
\begin{align}
 v^2(Q) = \frac{4\{m_Z^2  + \text{Re}[\Pi^{T}_{ZZ}(m_Z^2)]\} }
{g^{\prime 2}(Q) + g^2(Q)}~, 
\label{eq:runVEV}
\end{align}
where $\Pi^{T}_{ZZ}(m_Z^2)$ is the transverse part of the $Z$-boson
self-energy in the $\overline{\text{MS}}$ scheme with external momentum
set to be $p^2=m_Z^2$, and evaluated
at the renormalization scale $Q$.  

\begin{figure}[t!]
\begin{center}
\includegraphics[clip, width = 0.6 \textwidth]{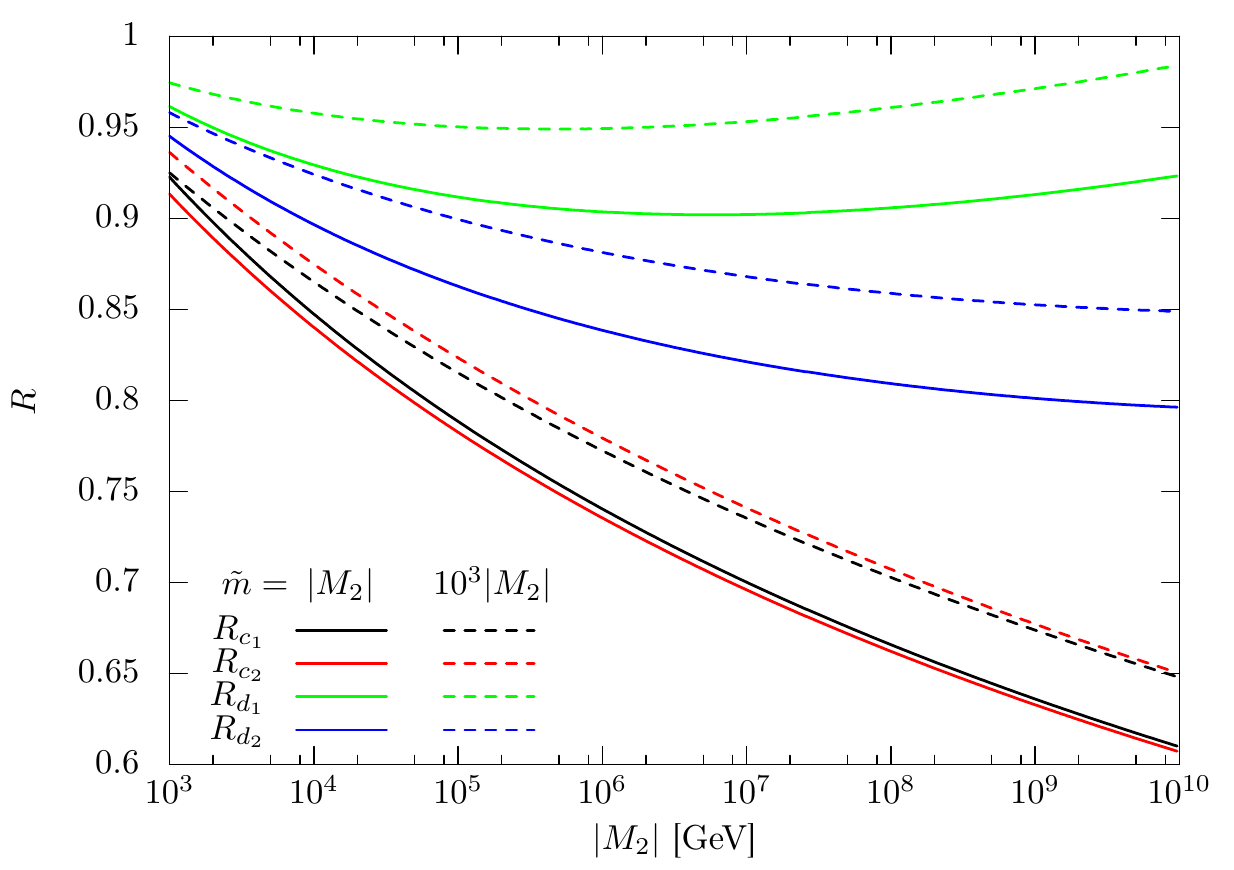}
\caption{
The ratios $R$'s as functions the Wino mass.
We set $\tan\beta = 2$, and $\mu =100$~GeV. Black, red, green, blue
 lines correspond to $R_{c_1}$, $R_{c_2}$, $R_{d_1}$, and $R_{d_2}$,
 respectively. Solid lines are for $\widetilde{m}=|M_2|$, while dashed
 lines for $\widetilde{m}=10^3|M_2|$.
}
\label{fig:ex1}
\end{center}
\end{figure}

In Fig. \ref{fig:ex1}, we show the ratios $R_{c_i}$ and $R_{d_i}$
$(i=1,2)$ as functions of the Wino mass $|M_2|$. Here we assume $\tan\beta
= 2$ and $\mu=100$~GeV. The black, red, green, blue
 lines correspond to $R_{c_1}$, $R_{c_2}$, $R_{d_1}$, and $R_{d_2}$,
 respectively. In solid lines, we take the SUSY breaking scale
 $\widetilde{m}$ to be $\widetilde{m}=|M_2|$, while in dashed
 lines $\widetilde{m}=10^3|M_2|$. From this figure, we find that the
 renormalization group effects 
modify the Wilson coefficients by $\mathcal{O}(10)$\%. The difference is
particularly important when one considers the mass difference in the
Higgsino components, as we will see below. Moreover, the figure shows
that the results depend not only on the Higgsino and gaugino masses, but
also on the SUSY breaking scale $\widetilde{m}$. This is because the
Higgsino-gaugino Yukawa couplings run differently from the gauge
couplings below the SUSY breaking scale
\cite{ArkaniHamed:2004fb,*Giudice:2004tc,*ArkaniHamed:2004yi},\footnote{The
RGEs of the Higgsino-gaugino couplings are given in
Appendix~\ref{sec:RGEs}. In addition, we have included finite threshold
corrections at the SUSY breaking scale. } and accordingly the relations
\eqref{eq:gauginocoupmatch} do not hold at the gaugino mass scale. This
then affects the ratios $R_{c_i}$ and $R_{d_i}$, especially when the
SUSY breaking scale is much higher than the gaugino mass scale.


\section{Higgsino Dark Matter Search}
\label{sec:DMsearch}

As mentioned in the Introduction, the neutral Higgsino LSP with a mass
of around TeV scale can be a dark matter candidate. In fact, the thermal
relic abundance of the Higgsino LSP is consistent with the observed DM
density when it has $\sim 1$~TeV mass \cite{Cirelli:2007xd}. In this
section, we assume that the Higgsino LSP occupies the dominant component
of the DM in the Universe, and consider the constraints on the DM from
the direct detection experiments.\footnote{As for the indirect search of
the Higgsino DM, a robust limit is given in
Ref.~\cite{Geringer-Sameth:2014qqa} based on the observations of  Milky
Way's dwarf galaxies by Fermi Gamma-ray Space Telescope. According to
the results, the current bound on the DM mass is $m_{\rm DM} \simgt
200-300$ GeV.} The mass of the Higgsino DM is assumed to be lower than
1~TeV to satisfy the environment selection requirement discussed in the
Introduction.

\subsection{Inelastic Scattering}

Without the dimension-five effective operators, the Higgsino DM forms a
Dirac fermion. In this case, the $Z$-boson exchange process induces the
vector-vector coupling between the DM and a nucleon. Due to the coupling,
the spin-independent (SI) scattering cross sections between the DM and
nucleons are so large that this Dirac Higgsino scenario turns out to
be already excluded by the direct detection experiments. However, thanks
to the higher dimensional operators, the neutral components of Higgsino
split into two Majorana fermions $\widetilde{\chi}^0_1$ and
$\widetilde{\chi}^0_2$ with the mass difference $\Delta m$ given in
Eq.~\eqref{eq:neutmassdiff}. Since a Majorana fermion does not have
vector interactions, the Majorana Higgsino DM can avoid the bound from the
direct detection experiments.

\begin{figure}[t]
\begin{center}
\includegraphics[ width = 0.25 \textwidth]{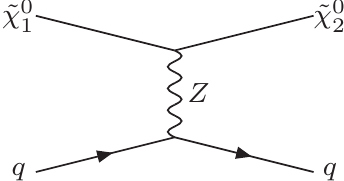}
\caption{Diagram which gives rise to the inelastic scattering process 
$\widetilde{\chi}^0_1 N\to \widetilde{\chi}^0_2 N$. }
\label{fig:ineldiagram}
\end{center}
\end{figure}

Nevertheless, if the mass difference $\Delta m$ is as small as
$\mathcal{O}(100)~\text{keV}$, inelastic scattering processes
$\widetilde{\chi}^0_1 N \to 
\widetilde{\chi}^0_2 N$ ($N$ denotes a nucleon) may occur through the
diagram in Fig.~\ref{fig:ineldiagram}. The inelastic scattering is also
restricted by the direct detection experiments, depending on the mass
difference \cite{TuckerSmith:2001hy,*TuckerSmith:2004jv}. Let us
consider the constraints on the mass difference 
$\Delta m$ by studying the process. This bound then can
be interpreted as an upper bound on the gaugino mass scale, as we will
see in what follows. 

By evaluating the diagram in Fig.~\ref{fig:ineldiagram}, we readily obtain
the effective Lagrangian for the vector-vector interaction between the
DM and quarks:
\begin{equation}
 {\cal L}_{\rm eff}=b_q \overline{\widetilde{\chi}^0_2}\gamma^\mu
  \widetilde{\chi}^0_1 \overline{q}\gamma_\mu q
+\text{h.c.} ~,
\end{equation}
with
\begin{equation}
 b_q=-\frac{iG_F}{\sqrt{2}}
(T^q_3 -2 Q_q \sin^2\theta_W)~,
\end{equation}
where $G_F$ is the Fermi constant, and $T^q_3$ and $Q_q$ are $+1/2$ and
$+2/3$ ($-1/2$ and $-1/3$) for up-type (down-type) quarks,
respectively. Since sea quarks and gluons cannot contribute to the
vector current, the effective vector couplings for proton and neutron
are readily obtained as the sum of the valence quark contributions.
By using the effective couplings,
we obtain the SI inelastic scattering cross section
of the Higgsino DM with a nucleus as
\begin{equation}
 \sigma_{\rm inelastic} =\frac{G_F^2}{8\pi}
[N-(1-4\sin^2\theta_W)Z]^2M_{\text{red}}^2~,
\end{equation}
where $M_{\text{red}}\equiv
m_{\widetilde{\chi}^0}m_T/(m_{\widetilde{\chi}^0}+m_T)$ is the reduced
mass in 
the DM-nucleus system with $m_T$ being the mass of the target nucleus,
and $Z$ and $N$ are the numbers of protons and neutrons in the
nucleus, respectively. In the case of the ${}^{131}$Xe target, for
example, $Z=54$ and $N=77$ with a mass of $m_T\sim 122$~GeV.

In a direct detection experiment, we search for the recoil energy $E_R$
of a target nucleus scattered off by the DM particle. The differential
scattering rate for the Higgsino DM is expressed as 
\begin{equation}
 \frac{dR}{dE_R}=\frac{N_Tm_T \rho_{\widetilde{\chi}^0_1}G_F^2}
{16\pi m_{\widetilde{\chi}^0}}
[N-(1-4\sin^2\theta_W)Z]^2 F^2(E_R)
\int_{v_{\text{min}}}^\infty \frac{f(v)}{v} dv~.
\label{eq:diffrate}
\end{equation}
Here, $N_T$ is the number of the target nuclei, $F^2(E_R)$ is a nuclear
form factor, 
$\rho_{\widetilde{\chi}^0_1}$ is the local DM density, and $f(v)$ is the
local DM velocity distribution. We use the same nuclear form factor as
that given in Ref.~\cite{Angle:2009xb} in the following calculation. The
DM density is assumed to be 
$\rho_{\widetilde{\chi}^0_1}= 0.3~\text{GeV}/\text{cm}^3$. For $f(v)$, we
use a Maxwell-Boltzmann velocity distribution with the escape velocity
$v_{\text{esc}}$, in which the circular speed of the Sun is assumed to
be $v_0=220~\text{km/s}$. For the choice of the astrophysical parameters
and the effects of their uncertainties on resultant constraints, see
Ref.~\cite{McCabe:2010zh}. In 
Eq.~\eqref{eq:diffrate}, the minimum speed $v_{\text{min}}$ is given by
\begin{equation}
 v_{\text{min}}=\frac{c}{\sqrt{2m_TE_R}} \biggl(
\frac{m_TE_R}{M_{\text{red}}}+\Delta m
\biggr)~.
\end{equation}
Dark matter direct detection experiments have good sensitivities for the
recoil energy $E_R$ smaller than $\mathcal{O}(100)$~keV. Thus, if the 
mass difference $\Delta m$ is also smaller than $\mathcal{O}(100)$~keV,
it significantly affects the direct detection rate. The effects enable
us to probe or constraint $\Delta m$ in the region. 

\begin{figure}[t!]
\begin{center}
\includegraphics[clip, width = 0.6 \textwidth]{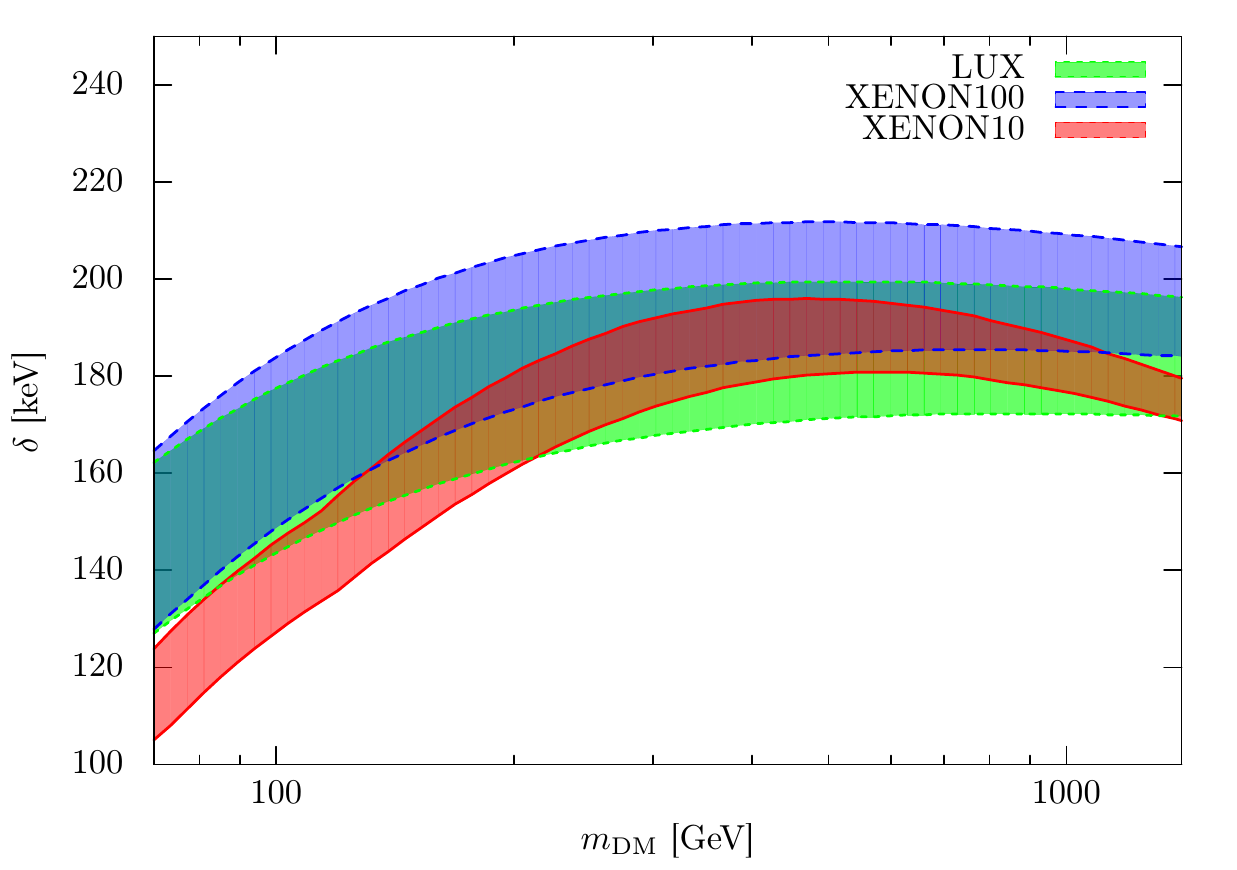}
\caption{
 Lower limits on $\Delta m$ at 90\% C.L as functions of DM mass
 $m_{\text{DM}}$. Red, blue, and green bands are computed based on the
 data provided by the XENON10
 \cite{Angle:2009xb}, XENON100 \cite{Aprile:2012nq}, and LUX
 \cite{Akerib:2013tjd} experiments, respectively. Upper (lower) line on
 each band corresponds to $v_{\text{esc}} =650$ (500)~km/s. 
}
\label{fig:inelbound}
\end{center}
\end{figure}

In Fig.~\ref{fig:inelbound}, we show the 90\% C.L. lower limits on
$\Delta m$ as functions of the DM mass $m_{\text{DM}}$. The red, blue,
and green bands show the constraints obtained from the data sets of the
XENON10 ($E_R<250$~keV) \cite{Angle:2009xb}, XENON100 ($E_R<50$~keV)
\cite{Aprile:2012nq}, and LUX ($E_R<36$~keV) \cite{Akerib:2013tjd}
experiments, respectively. The upper (lower) line on each band
corresponds to $v_{\text{esc}} =650$ (500)~km/s. To evaluate the limits,
we have used the $p_{\text{max}}$ method following
Ref.~\cite{Yellin:2002xd}. Slightly weaker limits are also provided in
the XENON10 \cite{Angle:2009xb}, CDMS~II \cite{Ahmed:2010hw}, and
XENON100 \cite{Aprile:2011ts} collaborations, though their analyses are
optimized to the parameter regions which may account for the modulation
observed by the DAMA/LIBRA experiment \cite{Bernabei:2008yi,
*Bernabei:2010mq}. We find that, although the constraints highly depend
on the astrophysical parameters such as the escape velocity
$v_{\text{esc}}$, the current direct detection experiments have
sensitivities to $\Delta m \lesssim (120-200)$~keV in the case of the
Higgsino DM scenario.

\begin{figure}[t!]
\begin{center}
\includegraphics[clip, width = 0.6 \textwidth]{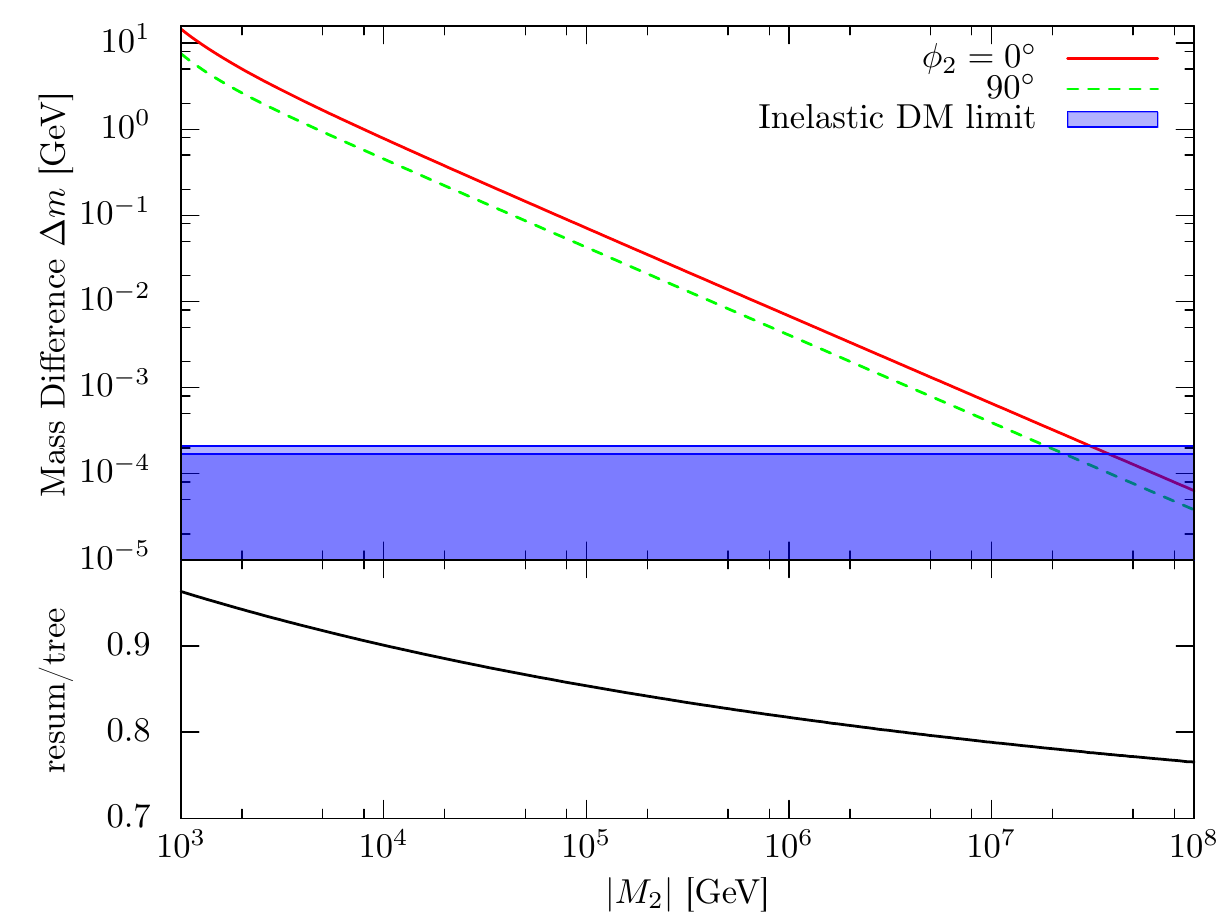}
\caption{Mass difference $\Delta m$ as functions of the Wino mass
 $|M_2|$ in solid lines. Here, we take $\tan \beta =2$, $\mu = 500$~GeV,
 $M_1=M_2$, and $\widetilde{m}=|M_2|$. Red-solid and green-dashed lines
 show the $\phi_2\equiv \arg{(M_2)}=0$ and $\pi/2$ cases,
 respectively. Dark (light) shaded region illustrates the weakest
 (strongest) bound given in Fig.~\ref{fig:inelbound}. The significance
 of the renormalization effects is shown in the lower graph. 
}
\label{fig:massdiff}
\end{center}
\end{figure}

Now we interpret the above constraints in terms of the bounds on the
gaugino mass scale. In the upper graph in Fig.~\ref{fig:massdiff}, we
plot the mass differences
$\Delta m$ as functions of the Wino mass $|M_2|$. Here, we take
$\tan \beta =2$, $\mu = +500$~GeV, $M_1=M_2$, and
$\widetilde{m}=|M_2|$. The red-solid and green-dashed lines show the
$\phi_2\equiv \arg{(M_2)}=0$ and $\pi/2$ cases, respectively. Results
for other phases lie between them. The dark (light) shaded
region illustrates the weakest (strongest) limits depicted in
Fig.~\ref{fig:inelbound}. The limits show that $M_2\gtrsim 4\times
10^4$~TeV has been already excluded. Further, to see the size of the
renormalization effects, we show in the lower graph the ratio of the
mass differences computed with and without the resummation. It is
found that to accurately extract the information on the gaugino mass
scale, as well as the CP-nature in the gaugino-Higgsino system, to
consider the renormalization effects is inevitable.

Before concluding this subsection, let us comment on the prospects of the
Higgsino DM search based on the inelastic scattering. Unlike the XENON10
experiment, the current analyses of the XENON100 and LUX experiments are
not optimized for the inelastic scattering. If the energy range analyzed
in the LUX experiment is extended to $E_R=250$~keV with keeping the
signal acceptance rate comparable to the present one, $\Delta m\sim 250$
(300)~keV can be constrained for $v_{\text{esc}}=500$ (650)~km/s and
$m_{\rm DM}=500$ GeV. We highly encourage such an analysis.

\subsection{Elastic Scattering}

In the presence of the higher-dimensional operators, the elastic
scattering also occurs via the exchange of the Higgs boson and the
$Z$-boson. The former gives rise to the SI scattering and the latter
induces the spin-dependent (SD) one. In this subsection, we study these
scattering processes. We will find that the SI scattering gives the
lower bound on the gaugino mass scale, while the SD scattering is
negligible.

\begin{figure}[t!]
\begin{center}
\subfigure[SI]
 {\includegraphics[clip, width = 0.25 \textwidth]{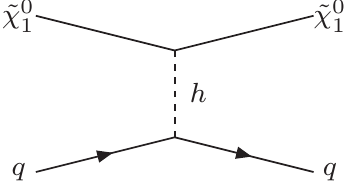}
 \label{fig:SIdiagram}}
\hspace{0.1\textwidth}
\subfigure[SD]
 {\includegraphics[clip, width = 0.25 \textwidth]{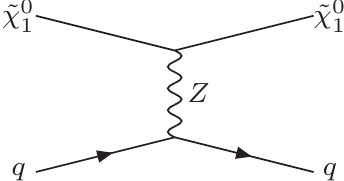}
 \label{fig:SDdiagram}}
\caption{
Diagrams induce the couplings of the Higgsino DM with quarks in the
 presence of the higher-dimensional operators. 
}
\label{fig:elasticfig}
\end{center}
\end{figure}

The SI effective
interactions between the DM and quarks/gluon are induced via the Higgs
exchange processes. The SI effective couplings of the DM with quarks
are generated by the diagram shown in
Fig.~\ref{fig:SIdiagram}. They are expressed in terms of the effective
operators as
\begin{equation}
\mathcal{L}_{\text{eff}}=\sum_{q}f_q\overline{\widetilde{\chi}^0_1}
\widetilde{\chi}^0_1 m_q \overline{q}q~,
\end{equation}
with 
\begin{equation}
 f_q=-\frac{1}{2m_h^2}{\rm Re}[c_1 e^{-i(\phi+\phi_\mu)}+ 
c_2 e^{i(\phi-\phi_\mu)}
+d_1e^{-i\phi_\mu}]~.
\end{equation}
Here, $m_h$ is the mass of the Higgs boson.
From the expression, we find that the SI interactions depend on the CP
phases in the Higgsino mass and the Wilson coefficients. With the
coupling $f_q$, the Higgsino DM-nucleon effective coupling $f_N$ is
written as
\begin{equation}
 \frac{f_N}{m_N}=\sum_{q=u,d,s}f_q f_{T_q}^{(N)} 
+\frac{2}{27} \sum_{Q=c,b,t}f_Q f_{TG}^{(N)}~,
\end{equation} 
where $f^{(p)}_{Tu}=0.019$, $f^{(p)}_{Td}=0.027$, $f^{(p)}_{Ts}=0.009$
for proton and $f^{(n)}_{Tu}=0.013$, $f^{(n)}_{Td}=0.040$,
$f^{(n)}_{Ts}=0.009$ for neutron, and $f^{(N)}_{TG}\equiv
1-\sum_{q=u,d,s}f^{(N)}_{T_q}$. They are computed from the recent
results of the lattice QCD simulations \cite{Young:2009zb,
Oksuzian:2012rzb}. The SI elastic scattering cross section of the
Higgsino DM with a target nucleus is then given as follows:
\begin{equation}
 \sigma_{\text{SI}}=\frac{4}{\pi}M_{\text{red}}^2
(Zf_p+Nf_n)^2~.
\end{equation}

In addition to the contribution, there exists the electroweak gauge
boson contribution at loop-level. The contribution is presented in
Ref.~\cite{Hisano:2011cs, *Hisano:2012wm}, and we take it into account
in the following analysis.

The SD scattering is, on the other hand, induced by the $Z$-boson
exchange process illustrated in Fig.~\ref{fig:SDdiagram}. The
interactions are expressed in terms of the following effective
Lagrangian: 
\begin{equation}
  {\cal L}_{\rm eff}=d_q \overline{\widetilde{\chi}^0_1}\gamma^\mu
 \gamma_5 \widetilde{\chi}^0_1 \overline{q}\gamma_\mu \gamma_5q ~.
\end{equation}
By evaluating the diagram, we obtain
\begin{equation}
 d_q=\frac{G_F}{\sqrt{2}}\cos 2\theta T^q_3 ~.
\end{equation}
Since the coupling is suppressed by $\cos 2\theta$, and since the
current experimental limits on the SD scattering are much
weaker than those on the SI one, we can safely
neglect the contribution in our scenario. 

\begin{figure}[t!]
\begin{center}
\includegraphics[clip, width = 0.6 \textwidth]{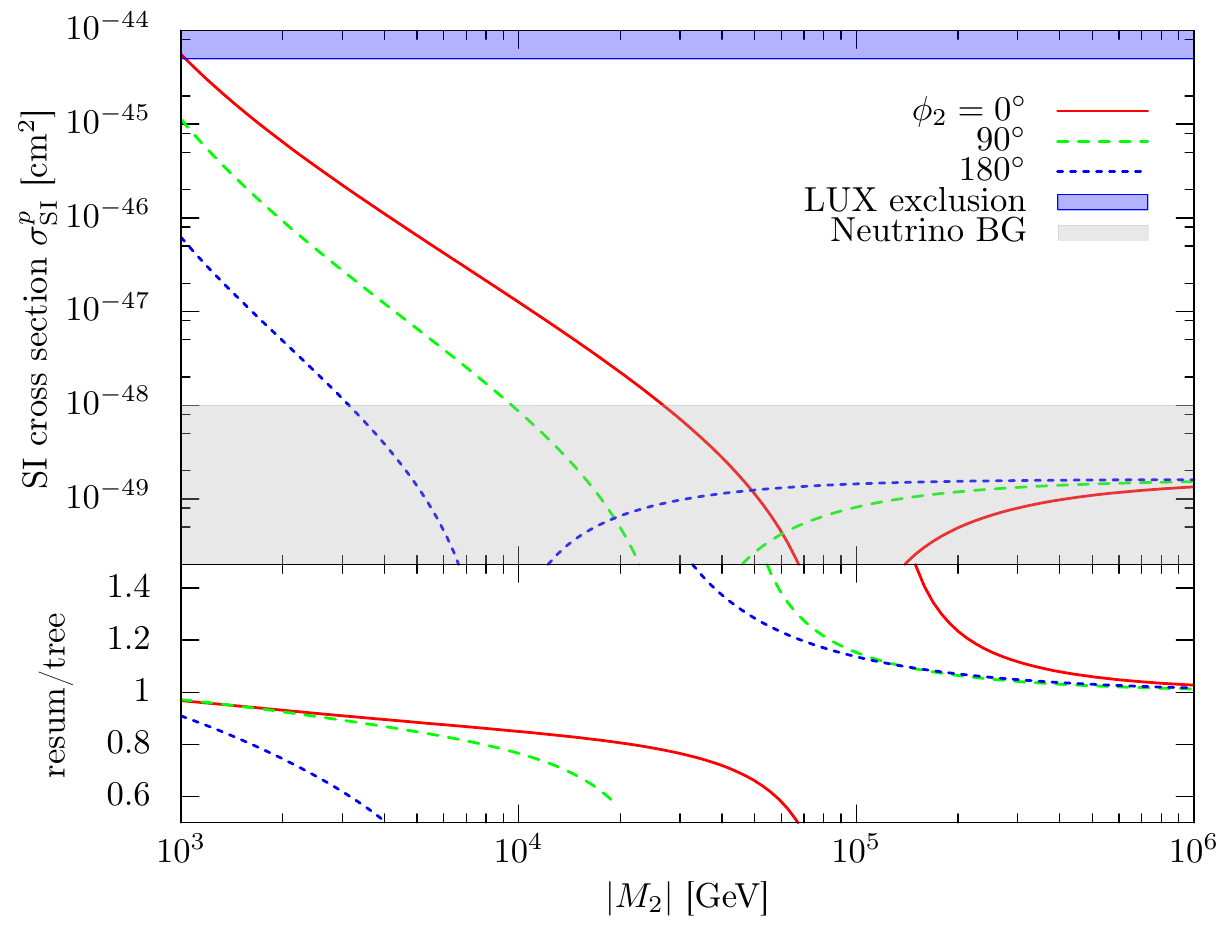}
\caption{
 SI scattering cross sections of the Higgsino DM with a proton as
 functions of $|M_2|$ in solid lines. Here we take $\tan\beta=2$, $\mu
 =500$~GeV, $M_1=M_2$ and $\widetilde{m}=|M_2|$. Red-solid, green-dashed,
 and blue short-dashed lines correspond to $\phi_2=\arg(M_2)=0$, $\pi/2$
 and $\pi$, respectively. Upper blue-shaded region is excluded by the
 LUX experiment \cite{Akerib:2013tjd}. Lower gray-shaded region
 represents the limitation of the direct detection experiments due to
 the neutrino background \cite{Billard:2013qya}. Lower panel represents
 the effects of the resummation on the calculation. 
}
\label{fig:elasticDM}
\end{center}
\end{figure}

Figure~\ref{fig:elasticDM} shows the SI scattering cross sections of the
Higgsino DM with a proton as functions of $|M_2|$ in solid lines. Here
we take $\tan\beta=2$, $\mu =500$~GeV, $M_1=M_2$ and
$\widetilde{m}=|M_2|$. The $\phi_2=\arg(M_2)=0$, $\pi/2$ and $\pi$, cases
are given in red-solid, green-dashed, and blue short-dashed lines,
respectively, and another choice of the CP-phase falls between them. The
upper blue-shaded region is already excluded by the LUX experiment
\cite{Akerib:2013tjd}. The lower gray-shaded region represents the
limitation of the direct detection 
experiments; once the experiments achieve the sensitivities to the cross
sections they will suffer from the neutrino background and cannot
distinguish the DM signal by means of the present
technique \cite{Billard:2013qya}. In addition, we show the effects of
the resummation on the calculation in the lower panel. 
As seen from the figure, the SI
scattering cross sections highly depend on the CP-phase in the
Higgsino-gaugino sector. When the gaugino scale is low enough,
the future direct detection experiments may detect the signal of the
DM. In higher gaugino mass regions, the electroweak loop effects
dominate the contribution to the SI scattering cross sections and the
resultant scattering cross sections become constant, though they are
much lower than the neutrino background limit.

\section{Electric Dipole Moments}
\label{sec:EDM}

\begin{figure}[t]
\begin{center}
\includegraphics[height=50mm]{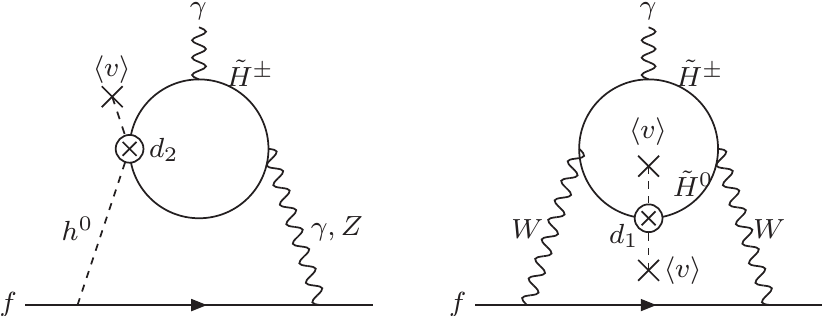}
\caption{Two-loop Barr-Zee diagrams which give rise to the EDMs. }
\label{fig:BZ}
\end{center}
\end{figure}
Generally, the MSSM induces new sources of CP violations, which may lead
to large electric dipole moments (EDMs) of the SM fermions. One of the
important contributions comes from one-loop diagrams which includes SUSY
scalar particles. Another significant contribution is two-loop diagrams
without the SUSY scalar particles. In the present ``lonely Higgsino''
scenario (typically when $\widetilde{m}\gg 10$ TeV), the latter
contribution is dominant.

As we noted above, the mass difference between the neutral components
depends on the new CP phases in the effective interactions in
Eqs.~\eqref{eq:effc} and \eqref{eq:effd}, and their effects can be
probed with the EDMs. The dominant
contribution to the EDMs comes from the two-loop Barr-Zee diagrams
\cite{Barr:1990vd} shown in Fig.~\ref{fig:BZ} \cite{Chang:2005ac,
Deshpande:2005gi, Giudice:2005rz}. To evaluate the contribution, let us
first show the Higgs-charged Higgsino vertex:  
\begin{equation}
 \mathcal{L}_{\text{int}}
=
-\text{Re}(d_2)vh\overline{\widetilde{H}^+}\widetilde{H}^+
+ \text{Im}(d_2)vh\overline{\widetilde{H}^+}i\gamma_5
\widetilde{H}^+~,
\end{equation}
and the CP-odd part (the second term) is relevant to our calculation.

The definition of the EDMs of fermion $f$ is
\begin{equation}
 \mathcal{L}_{\text{EDM}}
=-\frac{i}{2}d_f\overline{f}\sigma^{\mu\nu}\gamma_5F_{\mu\nu}f~.
\end{equation}
We now evaluate the contribution of the diagrams in
Fig.~\ref{fig:BZ} to the EDM $d_f$. The result is
given as follows \cite{Giudice:2005rz}:
\begin{equation}
 d_f = d^{h\gamma}_{f}+
d^{hZ}_{f}+d^{WW}_{f}~,
\end{equation}
with
\begin{align}
 d^{h\gamma}_{f} &= \frac{4e^3Q_f m_f}{(4\pi)^4}
\text{Im}\biggl(\frac{d_2}{\mu}\biggr)f_0\biggl(\frac{|\mu|^2}{m_h^2}\biggr)
~, \\
 d^{hZ}_f &= \frac{eg^2m_f}{(4\pi)^4}(T_f^3-2Q_f \sin^2\theta_W)
(1-\tan^2\theta_W)
\text{Im}\biggl(\frac{d_2}{\mu}\biggr)
f_1\biggl(\frac{m_Z^2}{m_h^2}, \frac{|\mu|^2}{m_h^2}\biggr)~, 
\label{eq:dhz}
\\
d_f^{WW}&=-\frac{eg^2m_f T^3_f}{(4\pi)^4}
\text{Im}\biggl(\frac{d_1+d_2}{\mu}\biggr)
f_0\biggl(\frac{|\mu|^2}{m_W^2}\biggr)~,
\end{align}
where $Q_f$, $T_f$ and $m_f$ are the electric charge, isospin and mass
of the fermion $f$, respectively, and $e$ is the electric charge of
positron. The loop functions are given by\footnote{
Here, we also give the analytic expression of $f_0(r)$ for
convenience: 
\begin{equation}
 f_0(r)=\frac{2r}{\sqrt{1-4r}}
\biggl[
\ln(r)\ln\biggl(\frac{\sqrt{1-4r}-1}{\sqrt{1-4r}+1}\biggr)
+\text{Li}_2\biggl(\frac{2}{1-\sqrt{1-4r}}\biggr)-
\text{Li}_2\biggl(\frac{2}{1+\sqrt{1-4r}}\biggr)
\biggr]~.
\end{equation}
}
\begin{align}
f_{0}(r) &= r \int_0^1 dx \frac{1}{r - x(1-x)}\ln\left(
\frac{r}{x(1-x)}
\right), \\[3pt]
f_{1}(r_1, r_2) &= \frac{1}{1-r_1}\biggl[
f_0(r_2)-r_1 f_0\biggl(\frac{r_2}{r_1}\biggr)
\biggr]~.
\end{align}

\begin{figure}[t!]
\begin{center}
\subfigure[Each contribution to the electron EDM. ]
{\includegraphics[clip, width = 0.42 \textwidth]{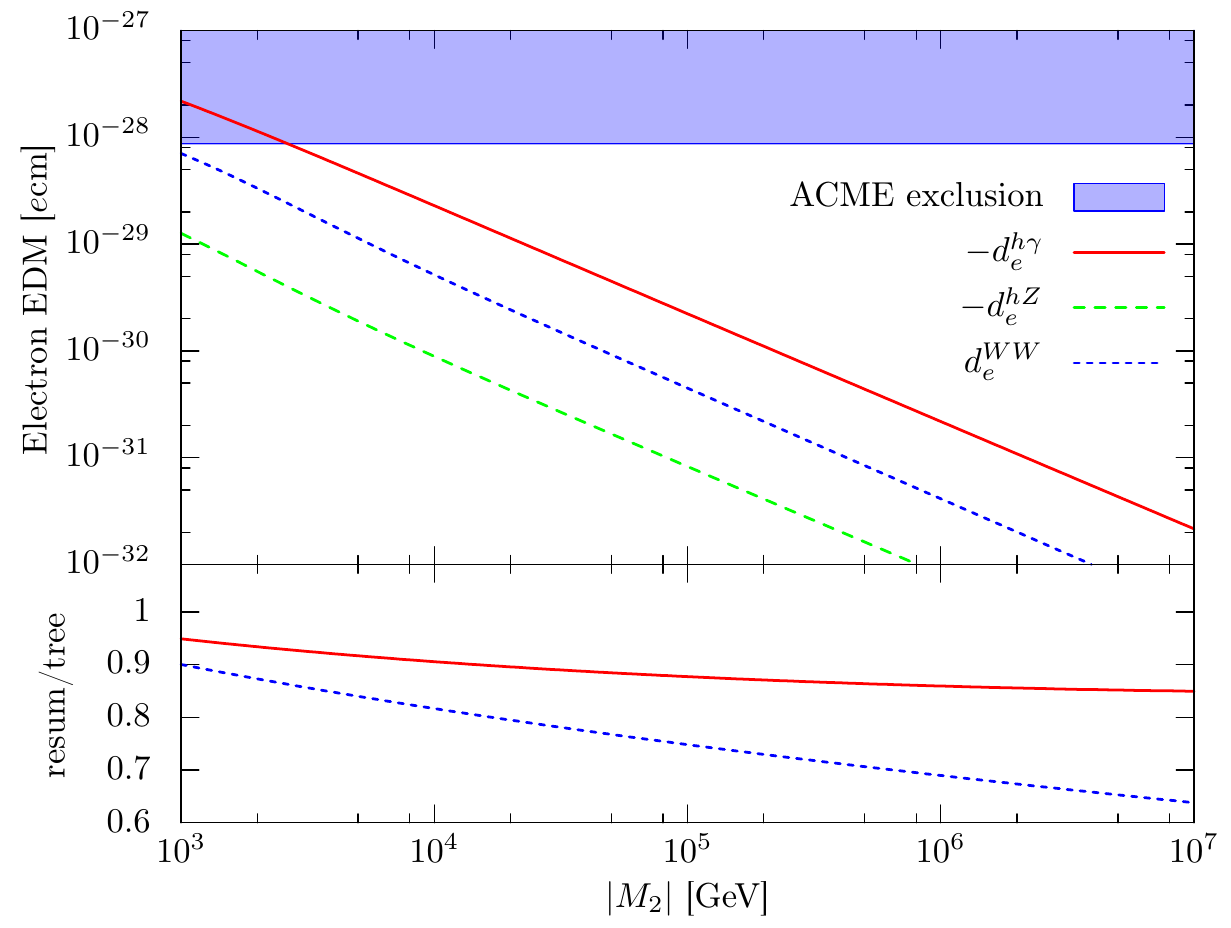}}
\hspace{0.08\textwidth}
\subfigure[Contour plot for the electron EDM.]
{\includegraphics[clip, width = 0.43 \textwidth]{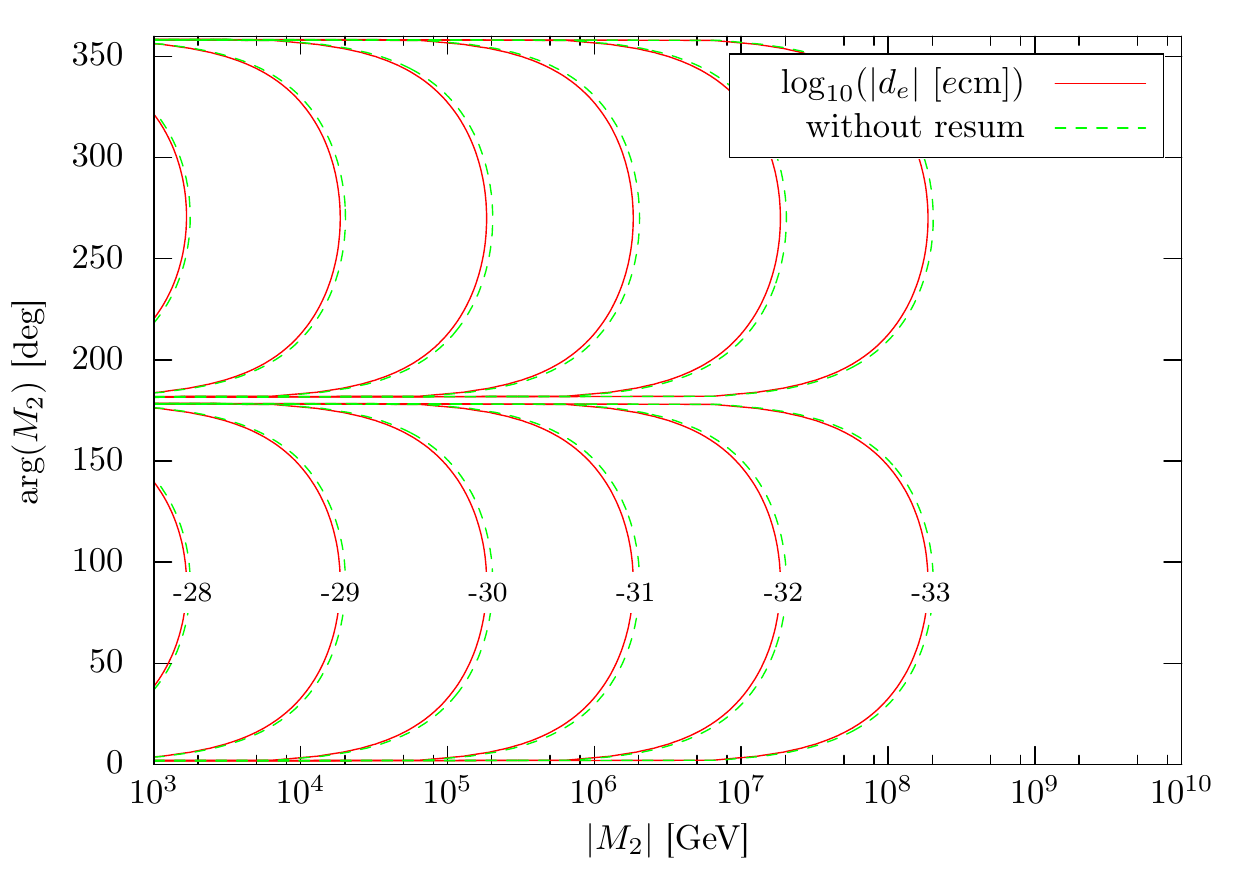}}
\caption{Results for the electron EDM. We take $\tan\beta=2$, $\mu
 =500$~GeV, $M_1=M_2$ and $\widetilde{m}=|M_2|$. Left: each contribution
 to the electron EDM as a function of $|M_2|$. Red-solid, blue short-dashed,
 and green-dashed lines show the contribution of $-d^{h\gamma}_e$,
 $d_e^{WW}$, and $-d_e^{hZ}$, respectively. We set
 $\phi_2=\pi/2$. Blue-shaded region is excluded by the ACME experiment
 \cite{Baron:2013eja}.  Lower
 panel illustrates the renormalization effects. Right: contour plot for
 the electron EDM. The red-solid and green-dashed lines represent the
 calculation with and without the resummation, respectively. 
}
\label{fig:edmfigs}
\end{center}
\end{figure}

By using the expressions, we evaluate the electron EDM, which gives the
most stringent bound on the Higgsino DM scenario at present. The results
are given in Fig.~\ref{fig:edmfigs}. In the left graph, we plot each
contribution to the electron EDM as a function of $|M_2|$. The red-solid,
blue short-dashed, and green-dashed lines show the contribution of
$-d^{h\gamma}_e$, $d_e^{WW}$, and $-d_e^{hZ}$, respectively. Here, we
take $\tan\beta=2$, $\mu =500$~GeV, $M_1=M_2$, $\phi_2=\pi/2$, and
$\widetilde{m}=|M_2|$. The blue-shaded region is excluded
by the current experimental limit given by the ACME
Collaboration \cite{Baron:2013eja}: $|d_e|<8.7\times 10^{-29}~e
\text{cm}$. The lower panel illustrates the
renormalization effects. It is found that the $\gamma$ and $Z$-boson
contributions have the opposite sign to the $W$-boson contribution.
The suppression of the $Z$-boson contribution results from a numerically
small factor of $T_e^3-2Q_e\sin^2 \theta_W=-(1-4\sin^2\theta_W)/2\simeq
-0.04$ in Eq.~\eqref{eq:dhz}. The total contribution is then shown in
the right panel as a contour plot. Here, the red-solid and green-dashed
lines represent the calculation with and without the resummation,
respectively. As can be seen from the figure, the present experiments
have sensitivities to well above the TeV regime, and has already
excluded a part of the parameter region shown in Fig.~\ref{fig:edmfigs}.

Future EDM experiments will have a few orders of magnitude improved 
sensitivity~\cite{Hudson:2011zz,Vutha:2009ux}, level of $d_e \sim
10^{-31}~e{\rm cm}$, or even smaller. In this case, the PeV scale
gauginos can be probed.

\section{Collider Signals}
\label{sec:collider}

\begin{figure}[t!]
\begin{center}
\includegraphics[clip, width = 0.6 \textwidth]{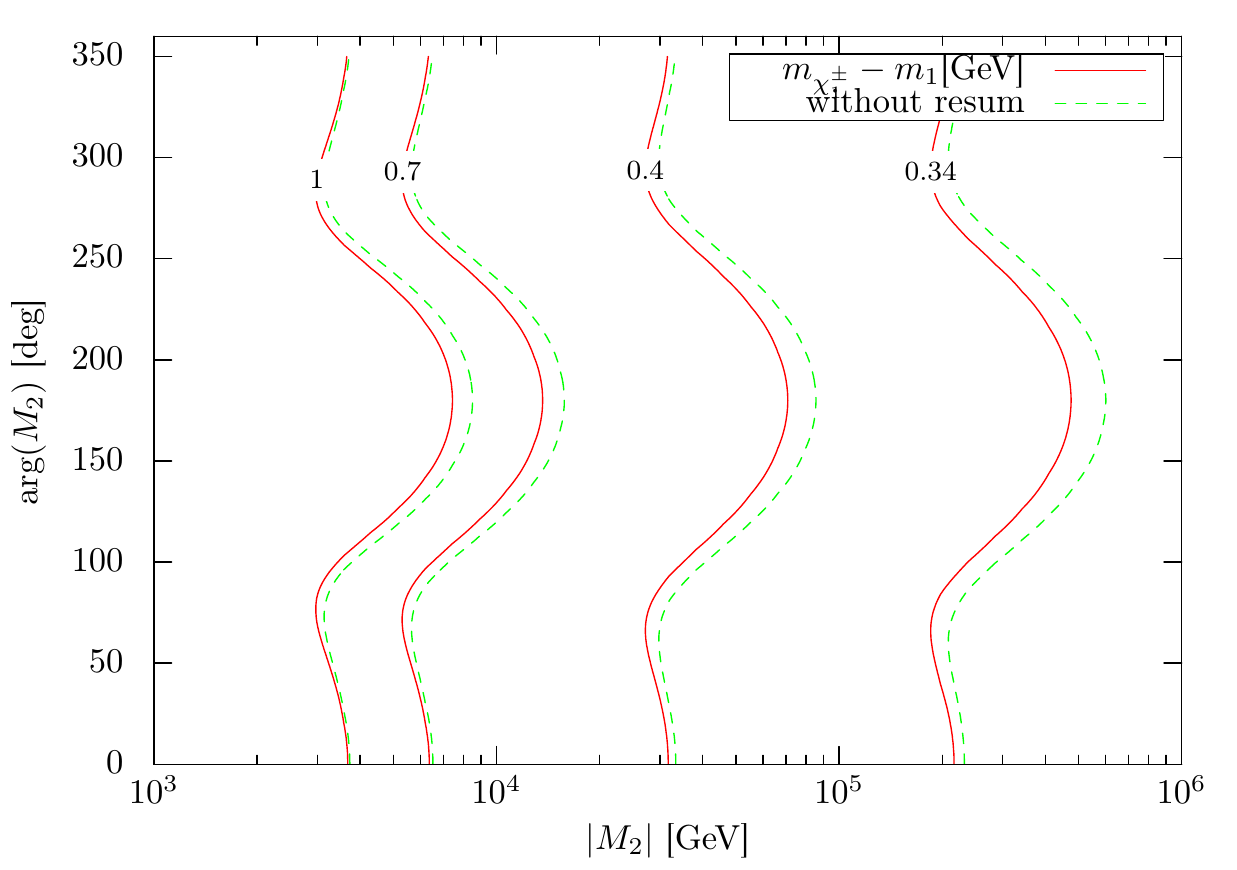}
\caption{Contour plot for the mass difference $\Delta m_+$ in the
 $\arg(M_2)-|M_2|$ plane. We take $\tan \beta =2$, $\mu = 500$~GeV,
 $M_1=M_2$, and $\widetilde{m}=|M_2|$. Red-solid and green-dashed lines
 show the calculations with and without the resummation effects,
 respectively. 
 }
\label{fig:massdiff_ch}
\end{center}
\end{figure}

As we have discussed above, the mass differences among the Higgsino-like
chargino and neutralinos $\Delta m$ and $\Delta m_+$ reflect the
high-scale SUSY breaking parameters. Therefore, detailed measurements of
the mass differences can reveal the high-energy physics. To that end, we
also need to perform theoretical calculations for the mass differences
accurately. The result for $\Delta m$ is already shown in
Fig.~\ref{fig:massdiff}. In Fig.~\ref{fig:massdiff_ch}, we show a
contour plot for the mass difference $\Delta m_+$ in the
$\arg(M_2)-|M_2|$ plane. Here, we take $\tan \beta =2$, $\mu = 500$~GeV,
$M_1=M_2$, and $\widetilde{m}=|M_2|$. Red-solid and green-dashed lines
show the calculations with and without the resummation effects,
respectively. We find that when $|M_2|=\mathcal{O}(1)$~TeV the
chargino-neutralino mass difference can be as large as
$\mathcal{O}(1)$~GeV. For heavier gaugino masses, on the other hand, the
mass difference approaches to a constant value. This is because in this
region the mass difference is determined by the electroweak loop
contribution in Eq.~\eqref{eq:cnrad}, and it reduces to $\Delta
m_+|_{\text{rad}} \simeq \alpha_2m_Z\sin^2\theta_W/2\simeq 350$~MeV in
the large gaugino mass limit as shown in Fig.~\ref{fig:cnrad}.

In the case of $\Delta m_+ \simgt m_{\pi}$, the chargino mainly decays
into hadrons and a neutralino. The decay length of the chargino is
\cite{Thomas:1998wy} 
\begin{align}
c\tau(\tilde \chi^\pm  \to \tilde \chi^0 \pi^\pm  ) = 1.1~{\rm cm} \left(
\frac{\Delta m_+}{300~{\rm MeV}}
 \right)^{-3} \left[ 1 - \frac{m^2_{\pi^\pm}}{\Delta m_+^2} \right]^{-1/2}.
\end{align}
In the case of the Higgsino LSP, $\Delta m_+ \simgt 300$~MeV, and thus
it is difficult to directly detect a charged track of the chargino,
unlike the Wino LSP case. In addition, smallness of the mass difference
makes it hard to even discover the Higgsino at a hadron collider
\cite{Han:2014kaa}.

However, at lepton colliders, it is possible to identify SUSY particle
production events by exploiting the hard photon tagging
\cite{Chen:1995yu}. With the process $e^+ e^- \to\tilde \chi^+ \tilde
\chi^- \gamma$, the LEP gives the lower limit on the chargino mass as
$m_{\chi^\pm}\simgt 90$~GeV \cite{Abbiendi:2002vz}. At a future lepton
collider, the measurement of the mass difference $\Delta m_+$ to an
accuracy of $O(1-10)$ \% is possible by observing the energy of the soft
pion from the $\chi^\pm$ decay for  $\Delta m_+ = O(100)~{\rm MeV} -
O(1)~{\rm GeV} $ \cite{Hensel:2002bu, Baer:2013cma, 
Berggren:2013vfa}. In this case, $\Delta m_+ |_{\rm tree}>O(10)$ MeV can
be discriminated. In other words, a few tens of TeV gauginos can be probed
by precisely measuring the chargino mass, as one can tell from
Fig.~\ref{fig:massdiff_ch}. In the analysis performed in
Fig.~\ref{fig:prcon} in the subsequent
section, we assume that a future lepton collider can determine the mass
difference of the chargino with an accuracy of 20\%
and show the corresponding gaugino mass scale that can be probed with
the mass measurements.

\section{Summary and Discussion}
\label{sec:summary}

\begin{figure}[t!]
\begin{center}
\subfigure[$\widetilde{m}=|M_2|$]
 {\includegraphics[clip, width = 0.42 \textwidth]{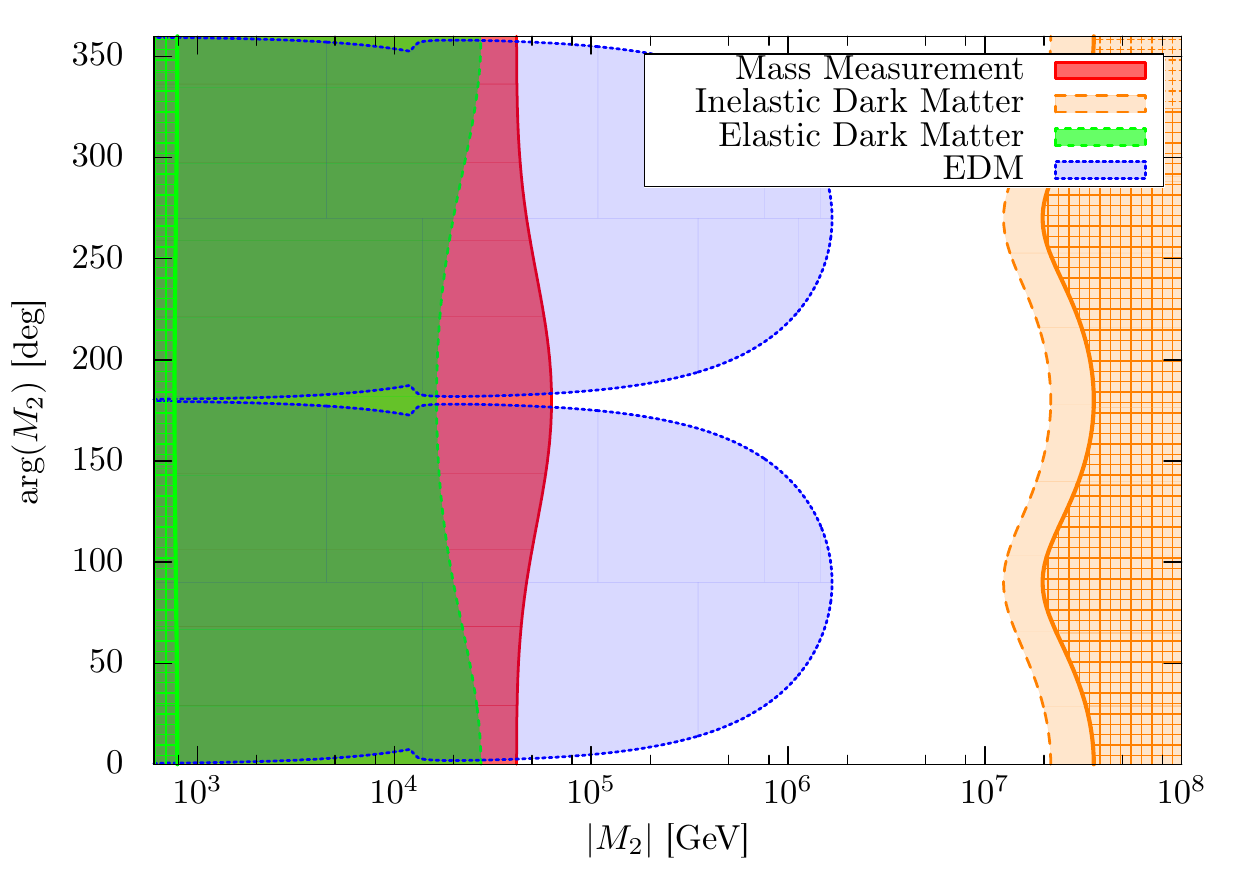}
 \label{fig:pc1}}
\hspace{0.1\textwidth}
\subfigure[$\widetilde{m}=10^2|M_2|$]
 {\includegraphics[clip, width = 0.42 \textwidth]{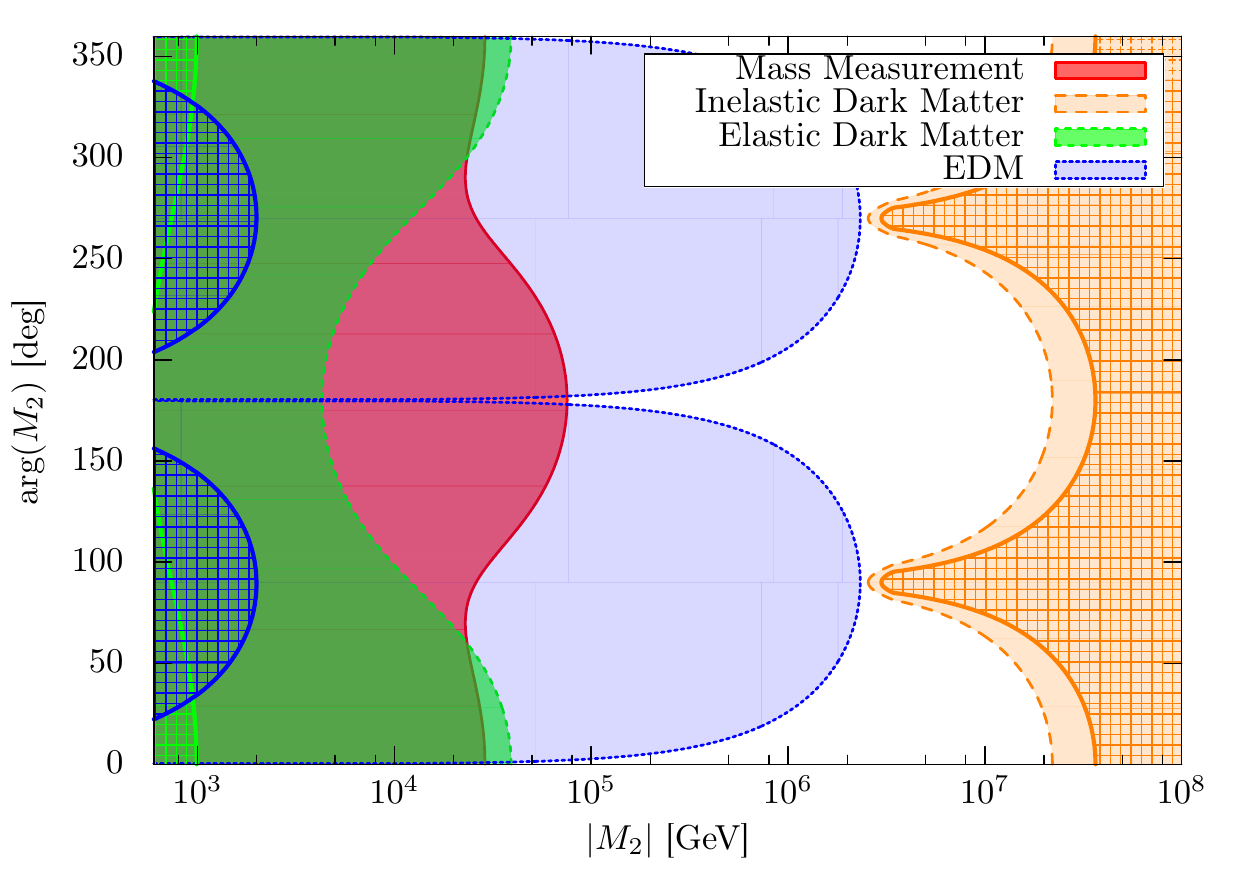}
 \label{fig:pc1e2}}
\caption{Current and future limits on the $|M_2|$-$\arg(M_2)$ space in
 the Higgsino DM scenario. Here, we set $\mu =+500$~GeV and $M_1=M_2=M_3$. 
 As for the future sensitivity, we assume $|d_e|>10^{-31}~e{\rm cm}$,
 $\sigma_{\rm SI} > 10^{-48}~{\rm cm}^2$, $\Delta m < 300$ keV and
 $\Delta m_+|_{\rm tree} > 0.2 \Delta m_+|_{\rm rad}$. 
}
\label{fig:prcon}
\end{center}
\end{figure}

Finally, we summarize the results which have been obtained so
far, and discuss the present constraints and future prospects on the
Higgsino DM scenario. The plots in Fig.~\ref{fig:prcon} show the
result. Here, we set $\mu =+500$~GeV, $M_1=M_2=M_3$ and $A$-terms are zero.
The left plot shows the case of $\widetilde{m}=|M_2|$, while the right
plot illustrates the $\widetilde{m}=10^2|M_2|$ case. The value of $\tan \beta$
is taken so that the Higgs mass is explained in the scenario. 
If an appropriate value of $\tan\beta \in [1:50]$ is not found, it is
set to be 1 (50) for the larger (smaller) Higgs mass. 
The mesh and shaded regions represent the present and future
constraints, respectively. For the EDM, we include only the Barr-Zee
contributions and omit the one-loop contribution with the sfermions in
the plots. It turns out that the future experiments have sensitivities
to probe a wide range of parameter regions and are complementary to each
other.

The heavier SUSY breaking scale can be also probed via measurement of
the spectrum of the cosmic gravitational background
\cite{Saito:2012bb}. This will give a good consistency check for the
MSSM.

Although we exploit a bottom-up approach to discuss the Higgsino DM
scenario in this paper, a top-down, or model-oriented approach is also
possible. If we consider a concrete model in which the Higgsino LSP
is realized, we may obtain some particular relations among the
parameters in the model. Such a relation sometimes affects the
nature of the Higgsino DM to a large extent. For example, let us
consider a high-scale SUSY model discussed in Ref.~\cite{Hall:2011jd}
where the Higgsino mass vanishes at tree level and is radiatively generated
via the gaugino-Higgs loop diagrams. In this case, the relative phase
between the Higgsino and gaugino mass terms is fixed: $\arg
(\mu/M_1)=\arg(\mu/M_2)= \pi$. Thus, the EDMs are not generated in the
scenario. Further, it turns out that the elastic scattering cross
sections are also significantly suppressed. The reason is the following. The
effective Higgsino-quark scalar coupling $f_q$ is given by
\begin{equation}
 f_q\simeq -\frac{g^2}{8m_h^2}\biggl(
\frac{\tan^2\theta_W}{M_1}+\frac{1}{M_2}
\biggr)(1-\sin 2\beta)~,
\end{equation}
with the gaugino masses taken to be real and positive. On the other
hand, to explain the mass of the Higgs boson in the scenario, $\tan\beta
\simeq 1$ is favored. As a result, the effective coupling, and therefore
the elastic scattering cross section as well, is extremely
suppressed. The bound coming from the inelastic scattering is also
evaded since the gaugino masses are $\mathcal{O}(10^{(2-3)})$~TeV to
realize a viable Higgsino DM. Consequently, the experimental constraints
on the scenario are significantly weakened.

In our work, we consider the effects of the SUSY particles on the
Higgsino DM properties based on the effective theoretical formalism. The
treatment is quite generic actually and applicable to other
high-energy theories or DM models. A straightforward generalization of
our study is to consider a generic multiplet of the
SU(2)$_L\otimes$U(1)$_L$ gauge group with its neutral component assumed
to be the DM---the so-called minimal DM scenario \cite{Cirelli:2005uq,
Cirelli:2007xd, Cirelli:2009uv}. The effects of the high-energy physics
on the DM are again described in terms of the higher-dimensional
operators, as discussed in Refs.~\cite{Hisano:2014kua,
Nagata:2014aoa}. In this scenario, the viable region for the DM mass
reaches as high as $\mathcal{O}(10)$~TeV. Thus, to thoroughly study the
possibilities, the precision experiments discussed in this paper play a
crucial role since it is much difficult to probe them in collider
searches. This highly motivates subsequent works in this direction.

\section*{Acknowledgments}

The work of N.N. is supported by Research Fellowships of the Japan Society
for the Promotion of Science for Young Scientists.

\section*{Appendix}
\appendix

\section{Diagonalization of a $2\times 2$ complex symmetric
 matrix}
\label{sec:takagi}

Here we give a set of formulae for the diagonalization of a $2\times 2$
complex symmetric matrix $M$ according to
Refs.~\cite{Takagi,Choi:2006fz}. Let us write the matrix as
\begin{equation}
 M=
\begin{pmatrix}
 a&c\\c&b
\end{pmatrix}
~,
\end{equation}
where $c\neq 0$ and $|a|\leq |b|$. We parametrize the $2\times 2$
unitary matrix $U$ by 
\begin{align}
 U=
\begin{pmatrix}
 e^{i\alpha}& 0 \\ 0& e^{i\beta}
\end{pmatrix}
\begin{pmatrix}
 \cos\theta & e^{-i\phi}\sin\theta \\
 -e^{i\phi} \sin\theta & \cos\theta
\end{pmatrix}
~,
\end{align}
which diagonalizes the matrix $M$ as
\begin{equation}
 U^*{\cal M}U^\dagger =
\begin{pmatrix}
 m_1&0\\ 0 & m_2
\end{pmatrix}
~,
\end{equation}
with $m_1$ and $m_2$ real and non-negative. Then, the above parameters
are given as follows:
\begin{align}
 m_{1,2}^2&=
\frac{1}{2}[|a|^2+|b|^2+2|c|^2\mp\sqrt{(|a|^2-|b|^2)^2
+4|a^*c+bc^*|^2}]~, \\
 \tan\theta &=\frac{|a|^2-|b|^2+\sqrt{(|a|^2-|b|^2)^2
+4|a^*c+bc^*|^2}}{2|a^*c+bc^*|}~, \\
  e^{i\phi}&=\frac{a^*c+bc^*}{|a^*c+bc^*|}~, \\
 \alpha &= \frac{1}{2}\arg\bigl(a-ce^{-i\phi}\tan\theta\bigr)~,\\
 \beta &= \frac{1}{2}\arg\bigl(b+ce^{i\phi}\tan\theta\bigr)~.
\end{align}

\section{Higgsino gauge interactions in the mass eigenbasis}

In this section, we list the gauge interactions of Higgsinos in the 
mass eigenbasis, for convenience. Here, we use the four-component
notation. The relevant interactions are given as follows: 
\begin{equation}
 {\cal L}_{\rm gauge}={\cal L}_W+{\cal L}_Z+{\cal L}_\gamma ~,
\end{equation}
with
\begin{align}
 {\cal
 L}_W=&-\frac{ge^{-\frac{i}{2}\phi}}{\sqrt{2}}
\overline{\widetilde{\chi}^+}\Slash{W}^+  
[e^{\frac{i}{2}(\alpha +\gamma)}\sin\theta P_L+e^{-\frac{i}{2}(\alpha +
 \gamma)}\cos\theta P_R] 
\widetilde{\chi}^0_1 \nonumber \\
&-\frac{ige^{-\frac{i}{2}\phi}}{\sqrt{2}}
\overline{\widetilde{\chi}^+}\Slash{W}^+  
[e^{\frac{i}{2}(\beta +\gamma)}\sin\theta P_L+e^{-\frac{i}{2}(\beta +
 \gamma)}\cos \theta P_R] 
\widetilde{\chi}^0_2 +\text{h.c.}~ , \\[3pt]
{\cal L}_Z=&+\frac{g_Z}{2}(1-2\sin ^2
 \theta_W)\overline{\widetilde{\chi}^+} \Slash{Z} \widetilde{\chi}^+ 
\nonumber \\
&+\frac{ig_Z}{4}[
\overline{\widetilde{\chi}^0_2}\Slash{Z}\widetilde{\chi}^0_1
-
\overline{\widetilde{\chi}^0_1}\Slash{Z}\widetilde{\chi}^0_2]
\nonumber \\
&+\frac{g_Z}{8}(\alpha-\beta)[
\overline{\widetilde{\chi}^0_2}\Slash{Z}\gamma_5\widetilde{\chi}^0_1
+
\overline{\widetilde{\chi}^0_1}\Slash{Z}\gamma_5\widetilde{\chi}^0_2]
\nonumber \\
&-\frac{g_Z}{4}\cos 2\theta [
\overline{\widetilde{\chi}^0_1}\Slash{Z}\gamma_5\widetilde{\chi}^0_1
-
\overline{\widetilde{\chi}^0_2}\Slash{Z}\gamma_5\widetilde{\chi}^0_2]
~, \\[3pt]
{\cal L}_\gamma =&-e \overline{\widetilde{\chi}^+} \Slash{A}
 \widetilde{\chi}^+ 
~,
\end{align}
where $P_{L/R}\equiv (1\mp \gamma_5)/2$ and $g_Z\equiv
\sqrt{g^2+g^{\prime 2}}$.

\section{Renormalization Group Equations}
\label{sec:RGEs}

Here, we present the RGEs other than those in the SM which are used in
the above calculation. First of all, the RGEs of the gauge couplings are
written as 
\begin{equation}
 \frac{dg_A}{d\ln Q}=\frac{b_Ag_A^3}{16\pi^2}~,
\end{equation}
where $g_1= g^\prime$, $g_2= g$, and $g_3=g_s$ is the strong gauge
coupling constant. Above the Higgsino threshold, the one-loop
beta-function coefficients $b_A$ are given by $(b_1,b_2,
b_3)=(\frac{15}{2}, -\frac{5}{2}, -7)$. After gauginos show up, we use
$(b_1,b_2, b_3)=(\frac{15}{2}, -\frac{7}{6}, -5)$.

Below the SUSY breaking scale, the running of the gaugino couplings
differs from that of the gauge couplings
\cite{ArkaniHamed:2004fb,*Giudice:2004tc,*ArkaniHamed:2004yi}. The RGEs
of the gaugino couplings $g_{iu}$ and $g_{id}$ $(i=1,2)$ in
Eq.~\eqref{eq:gauginohiggsinocoup} are
\begin{align}
 \frac{dg_{1u}}{d\ln Q}&=
\frac{1}{16\pi^2}\biggl[
g_{1u}\biggl(
\frac{3}{4}g_{1u}^2+\frac{3}{2}g_{1d}^2+\frac{3}{4}g_{2u}^2+3y_t^2
-\frac{3}{4}g^{\prime 2}-\frac{9}{4}g^2
\biggr)
+3g_{1d}g_{2u}g_{2d}
\biggr]~, \\
 \frac{dg_{1d}}{d\ln Q}&=
\frac{1}{16\pi^2}\biggl[
g_{1d}\biggl(
\frac{3}{4}g_{1d}^2+\frac{3}{2}g_{1u}^2+\frac{3}{4}g_{2d}^2+3y_t^2
-\frac{3}{4}g^{\prime 2}-\frac{9}{4}g^2
\biggr)
+3g_{1u}g_{2u}g_{2d}
\biggr]~, \\
 \frac{dg_{2u}}{d\ln Q}&=
\frac{1}{16\pi^2}\biggl[
g_{2u}\biggl(
\frac{5}{4}g_{2u}^2-\frac{1}{2}g_{2d}^2+\frac{1}{4}g_{1u}^2+3y_t^2
-\frac{3}{4}g^{\prime 2}-\frac{33}{4}g^2
\biggr)
+g_{2d}g_{1u}g_{1d}
\biggr]~, \\
 \frac{dg_{2d}}{d\ln Q}&=
\frac{1}{16\pi^2}\biggl[
g_{2d}\biggl(
\frac{5}{4}g_{2d}^2-\frac{1}{2}g_{2u}^2+\frac{1}{4}g_{1d}^2+3y_t^2
-\frac{3}{4}g^{\prime 2}-\frac{33}{4}g^2
\biggr)
+g_{2u}g_{1u}g_{1d}
\biggr]~,
\end{align}
while that of the top Yukawa coupling at one-loop level is given by
\begin{equation}
  \frac{d y_t}{d\ln Q}=\frac{1}{16\pi^2}
\biggl[\frac{9}{2}y_t^2-\frac{17}{12}g^{\prime
 2}-\frac{9}{4}g^2-8g_s^2+\frac{1}{2}(g_{1u}^2+g_{1d}^2)
+\frac{3}{2}(g_{2u}^2+g_{2d}^2)
\biggr]y_t ~.
\end{equation}

\bibliographystyle{aps}
\bibliography{ref}

\end{document}